\pdfoutput=1  % required for pdflatex compilation, otherwise errors with arXiv
\documentclass[5p,preprint]{elsarticle}

\graphicspath{ {./elsarticle-ecrc_CompFluid/} }

\usepackage{ecrc}

\volume{00}

\firstpage{1}

\journalname{Computers \& Fluids}

\runauth{D. Bartuschat et al.}

\jid{CompFluid}

\jnltitlelogo{CompFluid}

\usepackage{amssymb}
\CopyrightLine{2015}{Published by Elsevier Ltd.}

\usepackage{color}
\usepackage[T1]{fontenc}
\usepackage{lmodern}
\usepackage{amsmath}
\usepackage{amsfonts}
\usepackage{dsfont}
\usepackage{pdfpages}
\usepackage{epstopdf}
\usepackage{graphicx}
\usepackage[bookmarks=false]{hyperref}
\usepackage{float}
\usepackage{array}
\usepackage[hang]{subfigure} % \usepackage[hang,nooneline]{subfigure}
\usepackage{verbatim} % to somment multiple lines
\usepackage{enumitem} % Enumerated list with square brackets
\usepackage{xargs}                      % Use more than one optional parameter in a new commands
\usepackage[colorinlistoftodos,prependcaption,textsize=small]{todonotes}
\newcommandx{\Kat}[2][1=]{\todo[linecolor=SlateBlue3,backgroundcolor=SlateBlue3!25,bordercolor=SlateBlue3,#1]{#2}}
\usepackage{units}
\usepackage[mode=text,per-mode=symbol,exponent-product = \cdot]{siunitx}
\usepackage{multirow}
\usepackage{booktabs}   % required for cmidrule, etc.

\usepackage{lipsum} % to control vertical spacing between lines.

\definecolor{RYBred}{RGB}{225, 0,   0}
\definecolor{RYBblue}{RGB}{50, 120,   255}
\definecolor{RYBdarkblue}{RGB}{0, 0,140}
\definecolor{RYBSemiDarkRed}{RGB}{200, 0,0}
\definecolor{RYBDarkGreen}{RGB}{000, 130, 000}
\definecolor{RYBDarkCyan}{RGB}{000, 139, 139}
\definecolor{SlateGrey}{RGB}{112, 128, 144}
\definecolor{SlateBlue3}{RGB}{105, 089, 205}

\definecolor{darkgrey}{rgb}{0.3,0.3,0.3}
\definecolor{gray}{rgb}{0.4,0.4,0.4}

\newcommand{\fno}{{\sf m}}
\newcommand{\fnob}{{\sf l}}

\newcommand{\xb}{{\bf x}}
\newcommand{\xca}{\xb_{\fno}}
\newcommand{\pa}{{\bf t}_{\fno}}
\newcommand{\xcab}{\xb_{\fnob}}
\newcommand{\pab}{{\bf t}_{\fnob}}
\newcommand{\xcadot}{\dot{\xb}_{\fno}}
\newcommand{\padot}{\dot{\bf t}_{\fno}}
\newcommand{\xdot}{\dot{\xb}}
\newcommand{\tdot}{\dot{\bf t}}

\newcommand{\Fb}{{\bf F}}
\newcommand{\Mb}{{\bf M}}

\newcommand{\Gb}{{\bf G}}
\newcommand{\Sb}{{\bf S}}
\newcommand{\Db}{{\bf D}}
\newcommand{\fb}{{\bf f}}

\newcommand{\Ib}{{\bf I}}

\newcommand{\Vb}{{\bf V}}
\newcommand{\tb}{{\bf t}}

\newcommand{\Ucontrfno}{\Vb_{\fno}}

\newcommand{\Kb}{{\bf K}}
\newcommand{\Kbar}{\bar{\Kb}}
\newcommand{\Kbarl}[1]{\bar{\Kb} \left[ #1 \right] }

\newcommand{\ab}{{\bf a}}

\newcommand{\eeps}{\varepsilon}

\newcommand{\Rb}{{\bf R}}

\newcommand{\Rhat}{\hat{\Rb}}

\newcommand{\walberla}{\textsc{waLBerla}}
\newcommand{\Walberla}{\textsc{WaLBerla}}

\DeclareMathAlphabet{\mathpzc}{OT1}{pzc}{m}{it}
\newcommand{\pe}{$\mathpzc{pe}$}

\newcommand{\eg}{\mbox{e.\,g.}}
\newcommand{\ie}{\mbox{i.\,e.}}
\newcommand{\Ie}{\mbox{I.\,e.}}
\newcommand{\cf}{\mbox{cf.}}
\newcommand{\Fig}[1]{\mbox{Fig.\,\ref{#1}}}
\newcommand{\Sect}[1]{\mbox{Sec.\,\ref{#1}}}
\newcommand{\Tab}[1]{\mbox{Tab.\,\ref{#1}}}
\newcommand{\Eqn}[1]{\mbox{Eqn.\,(\ref{#1})}}

\begin{document}

\begin{frontmatter}



\dochead{}
%% e.g. \dochead{17th International Conference on Dynamical Processes in Excited States of Solids}

\title{Two Computational Models for Simulating the Tumbling Motion of Elongated Particles in Fluids}


% \author{Dominik Bartuschat, Ulrich R\"ude}
% \address{Lehrstuhl f\"ur Systemsimulation, Friedrich-Alexander Universit\"at Erlangen-N\"urnberg, Cauerstrasse 11, 91058 Erlangen, Germany}
% E-mail address: dominik.bartuschat@fau.de

\author[LSS]{Dominik Bartuschat\corref{cor1}}
\ead{dominik.bartuschat@cs.fau.de}
% \ead[url]{www10.cs.fau.de}
\cortext[cor1]{Corresponding author}

\author[TheorPhys]{Ellen Fischermeier}
\author[Nada]{Katarina Gustavsson}
\author[LSS]{Ulrich R\"ude}
% \author[TheorPhys]{Klaus Mecke}

\address[LSS]{Lehrstuhl f\"ur Systemsimulation, Friedrich-Alexander Universit\"at Erlangen-N\"urnberg, Cauerstrasse 11, 91058 Erlangen, Germany}
\address[TheorPhys]{Institut f\"ur Theoretische Physik I, FAU Erlangen-N\"urnberg, Staudtstrasse 7, 91058 Erlangen, Germany}
\address[Nada]{Department of Mathematics, Numerical Analysis/Linn\'e
  Flow Centre, Royal Institute of Technology (KTH), 100 44 Stockholm, Sweden}

%%%%%%%%%%%%%%%%%%%%%%%%%%%
%% Elsevier-specific end %%
%%%%%%%%%%%%%%%%%%%%%%%%%%%

% TODO Suggest three editors (Incl. Department, Institution, E-Mail, Reason)
\begin{abstract}
% Currently 135 words - no restriction found.
Suspensions with fiber-like particles in the low Reynolds number regime
are modeled by two different approaches
that both use a Lagrangian representation of individual particles.
The first method is the well-established formulation
based on Stokes flow that is formulated as 
integral equations.
It uses a slender body approximation for the fibers
to represent the interaction between them directly
without explicitly computing the flow field.
The second is a new technique
using the 3D lattice Boltzmann method
on parallel supercomputers.
Here the flow computation is coupled 
to a computational model of 
the dynamics of 
rigid bodies using fluid-structure interaction techniques.
Both methods can be applied to simulate fibers in fluid flow.
They are carefully validated and compared against
each other, exposing systematically their strengths and
weaknesses regarding their accuracy, the computational cost,
and possible model extensions.
% \comment
\end{abstract}


\end{frontmatter}

\section{Introduction}\label{sec:intro}
Flows with suspended solid phase occur in many applications, and thus
simulation techniques for such systems are receiving rapidly increasing interest.
An important and mathemactically interesting special case are fiber suspensions.
In this article we will 
investigate two different computational models that represent
the particulate solid phase in Lagrangian form.
Particles will be % modeled 
treated as rigid, elongated three-dimensional geometric objects.
Elongated is here understood as the situation that the shape of the particles
has one dimension significantly larger than the others,
and thus our two methods apply to suspensions with
fibers or rods.
For the fluid phase, we assume in this article creeping flows,
\ie\, the Reynolds number is small and the Stokes equation
can provide a sufficiently accurate approximation to the flow field.

The first method is based on a boundary integral formulation for Stokes flow,
and approximations exploiting the slenderness of the suspended particles.
This approach leads to an explicit representation of the hydrodynamic
interactions between particles that avoids computing the flow field
explicitly \cite{Poz92}. 
The resulting global system re\-presenting the hydrodynamic interactions
for this slender body formulation (SBF)
must be solved at each time step during a simulation~\cite{Jo80,ToGu06}.\\
The second method employs a full 3D Eulerian representation
of the fluid using the lattice Boltzmann method (LBM),
while the particles are represented as rigid, fully resolved, 
geometric objects that can move freely through the simulation domain. 
Here fluid-structure-interaction (FSI) mechanisms are used to couple the flow field to the
dynamics of the suspended particles.
This approach can lead to very
high computational cost
since the mesh for computing the flow field must be so fine that the geometry
of the suspended objects is resolved accurately enough.
Here using parallel supercomputers
is often inevitable~\cite{Goetz:2010:SC10}.

\begin{figure*}[h!t]
  \centering
  \subfigure{
  \includegraphics[width=0.145\textwidth,bb=0 0 451.937 942.869]{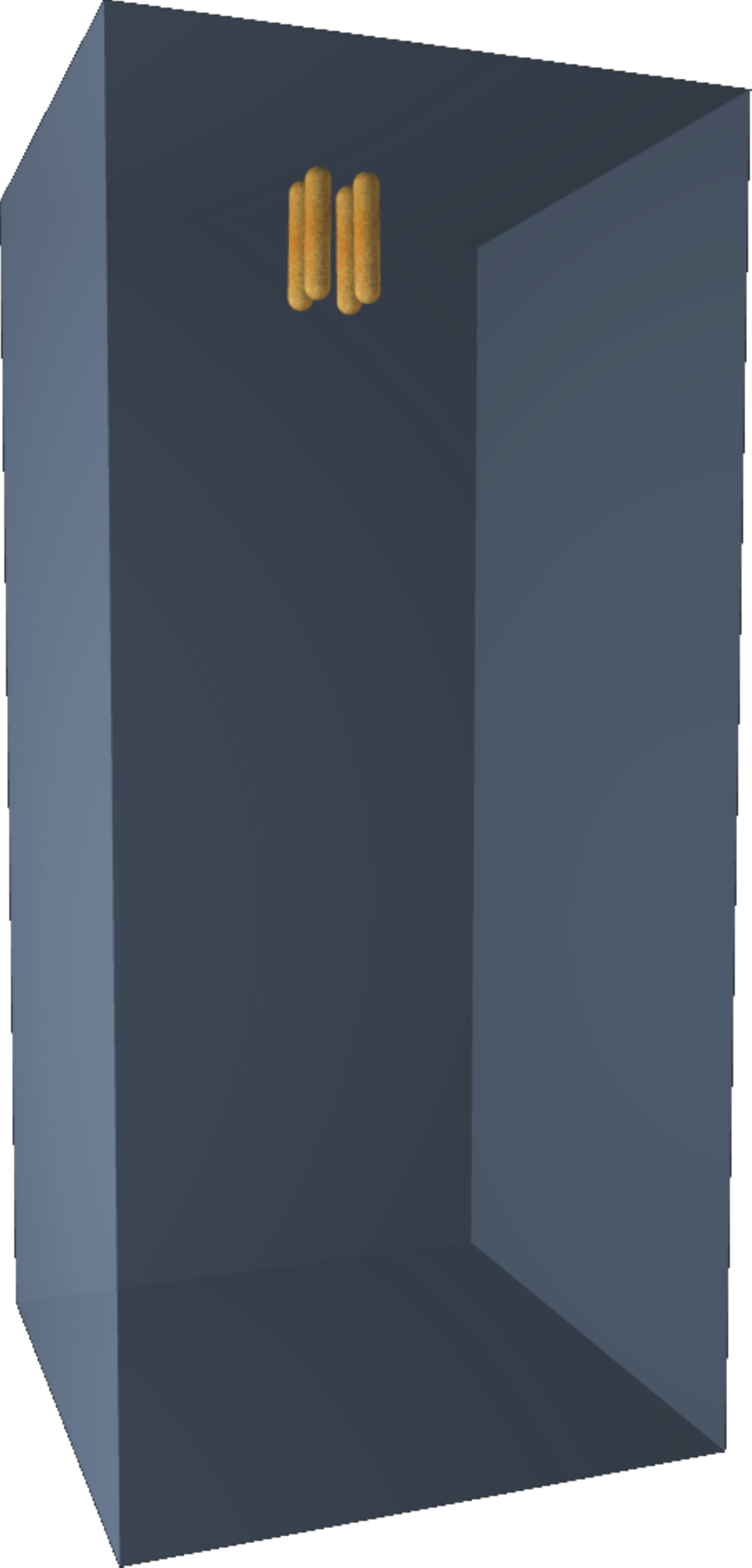} % bounding box required for arxiv, used MediaBox values in pdf file
  }%
  \subfigure{
  \includegraphics[width=0.145\textwidth,bb=0 0 451.937 942.869]{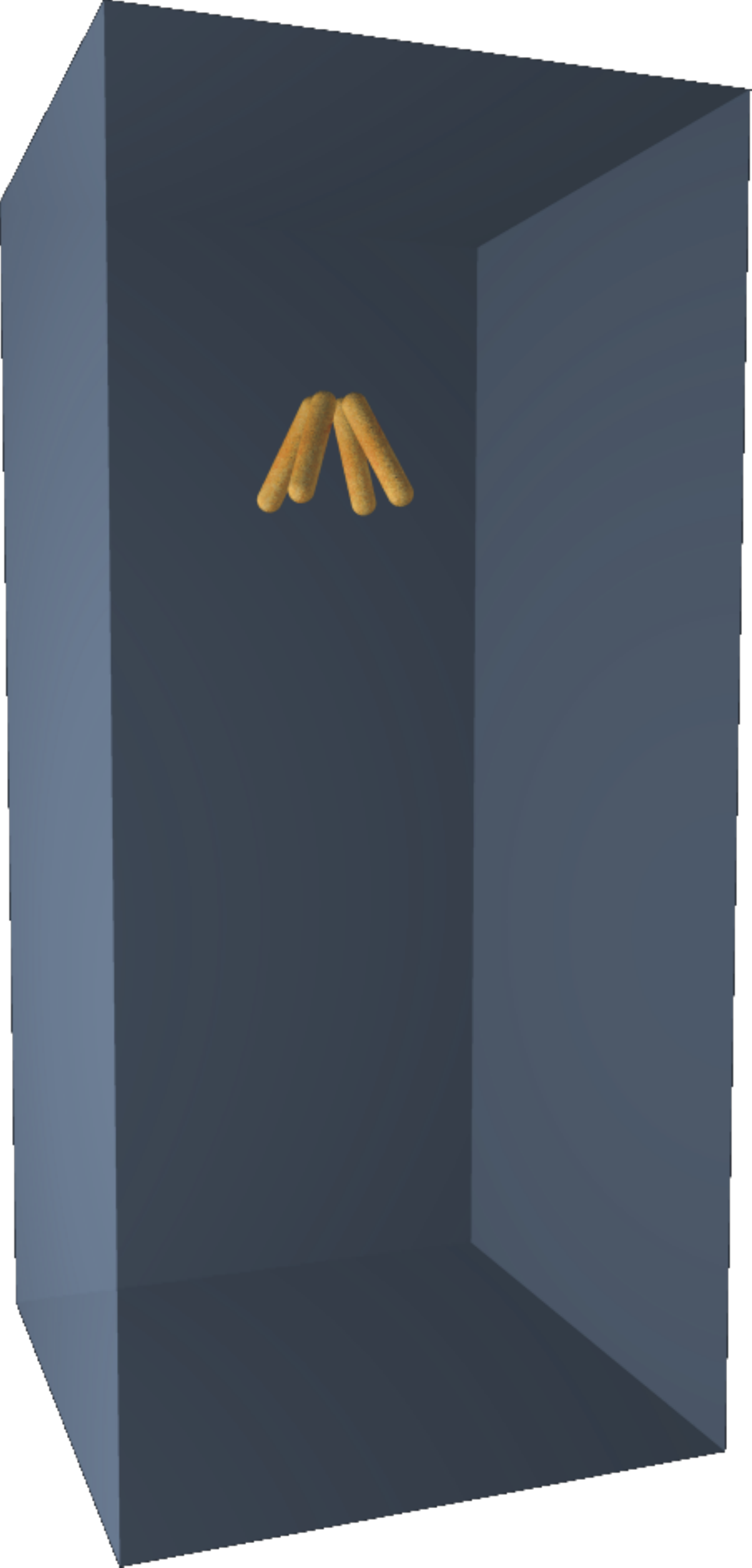}
  }%
  \subfigure{
  \includegraphics[width=0.145\textwidth,bb=0 0 451.937 942.869]{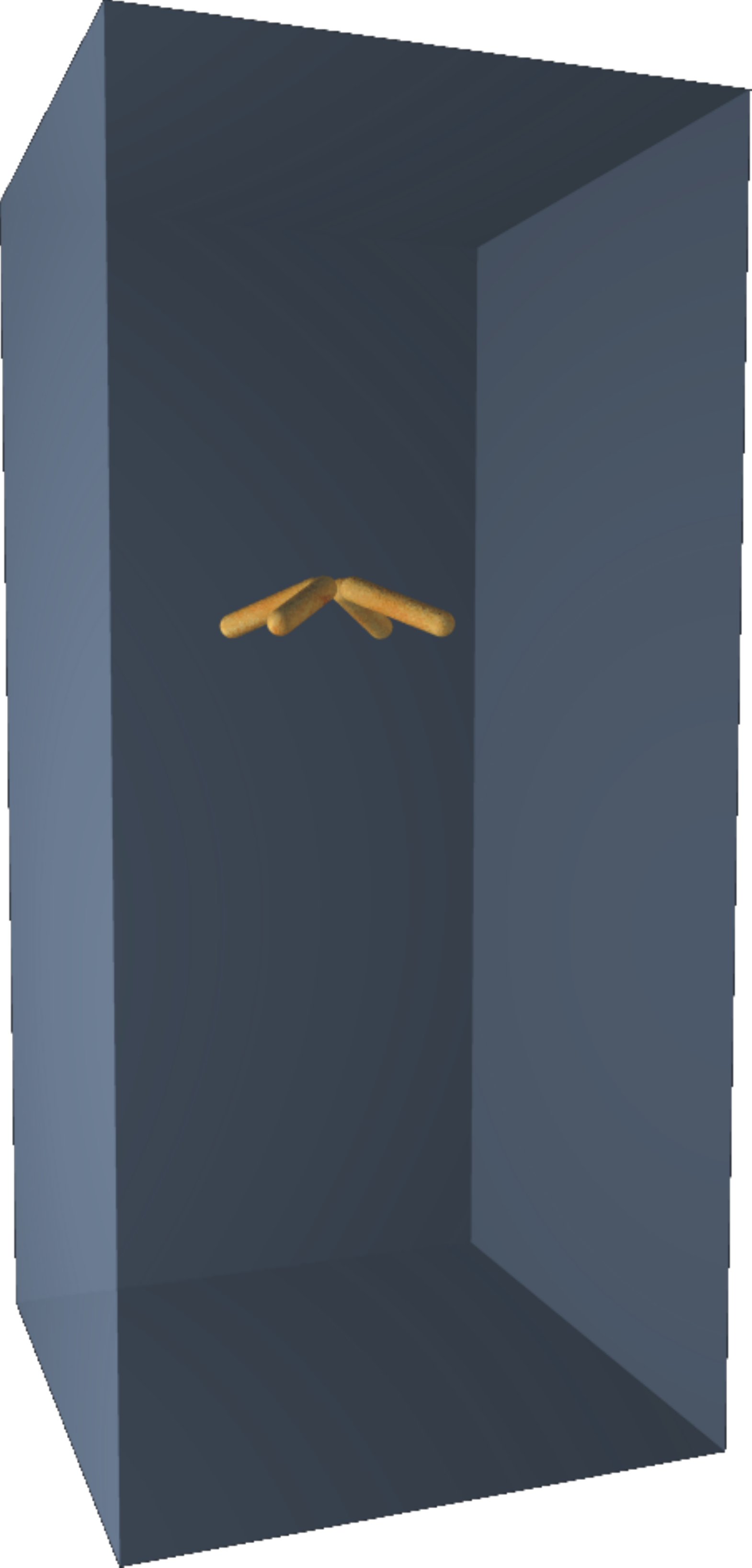}
  }%
  \subfigure{
  \includegraphics[width=0.145\textwidth,bb=0 0 451.937 942.869]{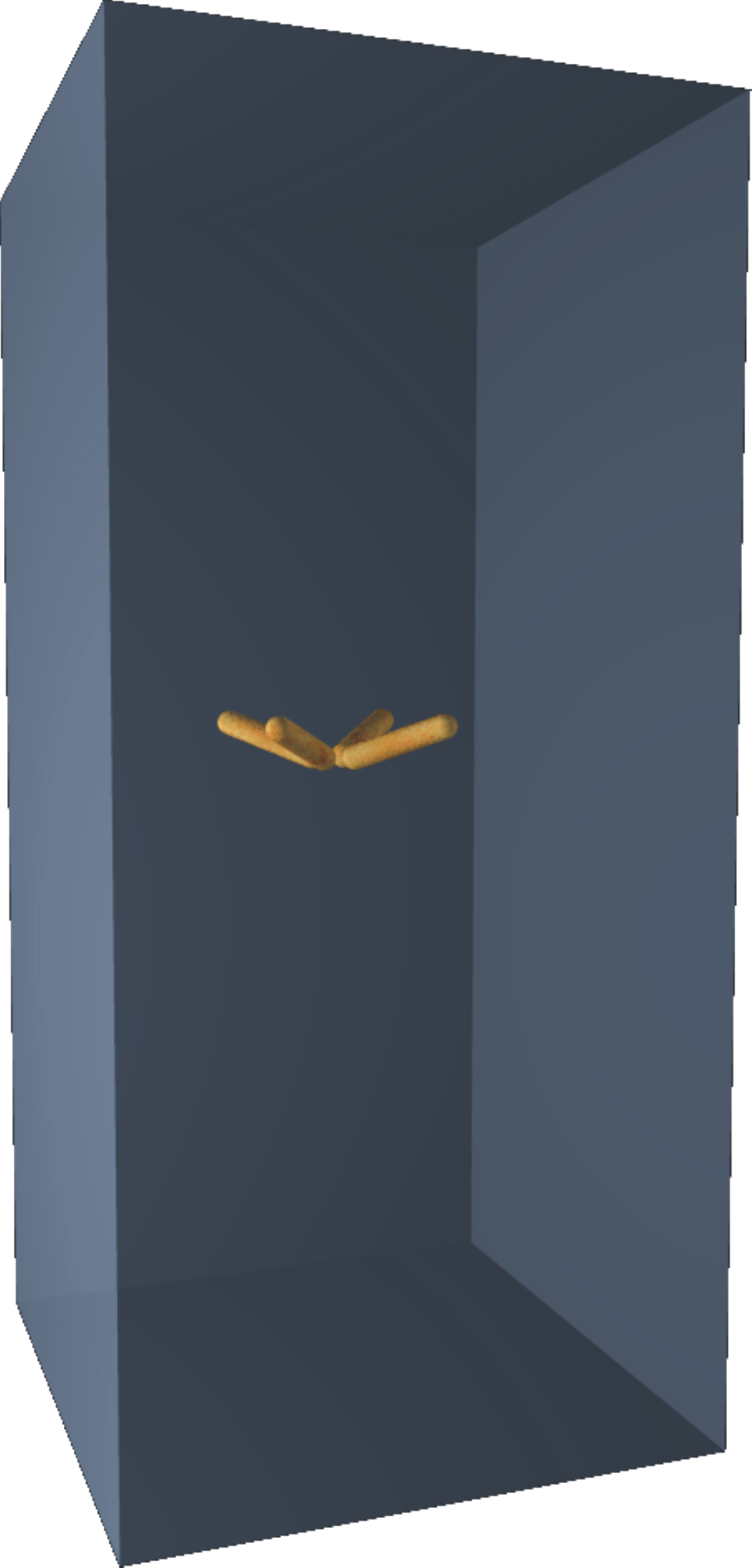}
  }%
  \subfigure{
  \includegraphics[width=0.145\textwidth,bb=0 0 451.937 942.869]{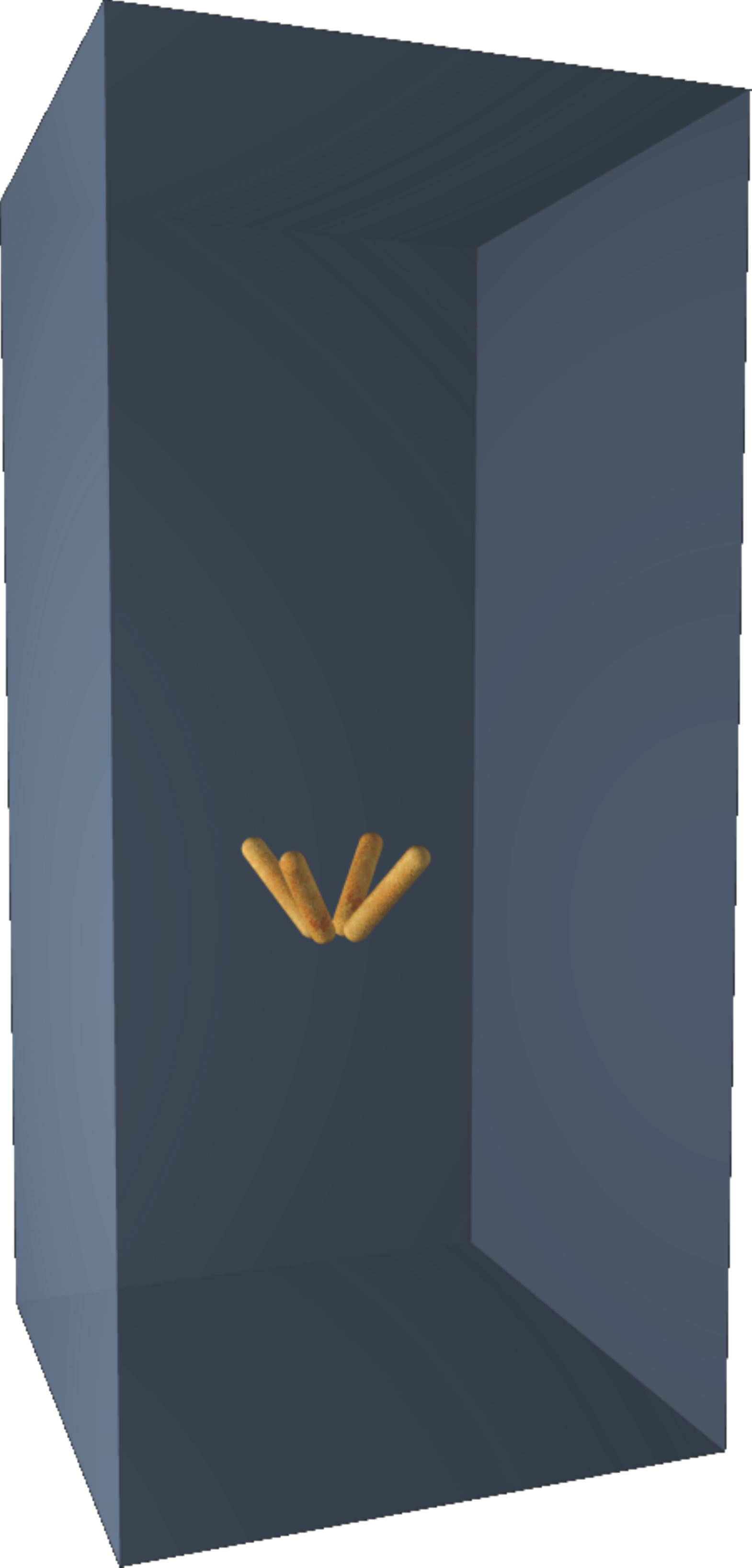}
  }
  \caption{Tumbling motion of four spherocylinders sedimenting in Stokes flow, simulated with LBM. \label{fig:TumblingCaps_FourCaps} }
\end{figure*}

Among the many interesting effects in fiber suspensions
we highlight here 
the tumbling trajectories of sedimenting fibers in Stokes flow, 
as visualized in the image sequence in~\Fig{fig:TumblingCaps_FourCaps}.
Experimental results for tumbling rods were presented
in~\cite{JungSedim2006PhysRevE.74.035302}.
Qualitatively correct simulations of this
dancing motion can be achieved both with the LBM and the SBF models.
To the best of our knowledge,
FSI-based simulations % of 
with the LBM that show this phenomenon
have not been reported before.

We will find that the SBF can be more efficient and
accurately represents the hydrodynamic interactions
when the assumptions on which it is based are well-satisfied.
The LBM, in contrast, can be used in more general settings
and permits many extensions, such as
simulating flows with higher Reynolds number and
more general boundary conditions,
and treating objects with different geometry than rigid fibers.
The LBM based approach can finally also represent scenarios
with particles colliding with each other or with the bounding walls.
These possible future extensions justify to develop
a method with higher computational cost.

Besides the development of the LBM with FSI techniques for 
fiber suspensions, the primary goal of this paper is the systematic comparison of
the two different simulation approaches
and to assess their strengths and weaknesses.
This comparison will also be used for the cross-validation of the
methods.
For complicated multi-physics scenarios such
as particles in suspension, the validation increasingly becomes a challenge in itself,
especially when analytical model solutions are not known,
and when only sparse data from physical experiments exist. 
In this case, the comparison of two different simulation
methodologies can serve as a powerful alternative
to assert the correctness of the models, the algorithms, and their implementation in software. 
In this paper, we study in particular the accuracy of the simulations
by comparing the translational settling velocity and the angular velocity 
to analytical solutions for the slender body motion.
As a more complex scenario with interacting particles, 
we will investigate tumbling fibers.
The simulation with four tumbling
fibers in \Fig{fig:TumblingCaps_FourCaps}
illustrates the potential of the methodology for
future applications that may involve many interacting particles.\footnote{Animation available via permalink:\\ \url{https://www10.cs.fau.de/permalink/eehaugh3oo}}
This article, however, 
is restricted to the
analysis and validation of the simpler case 
of two tumbling elongated particles. 

The SBF~\cite{Ba70,Jo80,ToGu06}
is an asymptotic method derived from the integral representation
of Stokes equations and leads to a model
consisting of a coupled system of one-dimensional integral
equations. The model uses a discrete representation of each ellipsoidal fiber in
terms of the fiber center line and takes into account the hydrodynamic
interactions of the fluid and the fibers. Due to the long-range nature
of the hydrodynamic interaction, a dense system of equations must be
solved.
This method is further described in \Sect{sec:slender}. 
The SBF has successfully been used for numerical
simulations of fiber suspension in gravity induced sedimentation, see
\eg{}~\cite{GuTo09, SaDaSh05}.

The LBM follows an alternative modeling paradigm
that explicitly represents the flow field in an Eulerian way, 
as described by the Stokes or Navier-Stokes equations.
A numerical simulation then requires the discretization on a grid, and the definition of interactions between particles and fluid and vice versa.
Such approaches have been studied \eg{} in~\cite{Glowinski:1999:Fictitious,prignitz2014particulate}.
Suspended particles must then be mapped to the grid similarly to an
immersed interface technique~\cite{Xu2006ImmersedInterf} or fictitious domain method~\cite{Glowinski:1999:Fictitious}.\\
In this article the LBM
is employed as an alternative to a Navier-Stokes solver.
The LBM uses a coupling between Eulerian fluid and
Lagrangian particles via the momentum exchange method~\cite{Ladd_1993_PartSuspPt1,nguyen2002lubrication,Bogner201571}
and by imposing moving boundary conditions on the fluid.
This approach allows the representation of arbitrary geometric shapes,
provided the grid resolution of the LBM method permits a sufficiently
accurate resolution.
For simulations of ensembles of several particles 
with good resolution, this leads to very large grids with small mesh size. 
These in turn lead to short time steps in the LBM algorithm. 
Combined, these effects can result in very large computational cost 
that can only be provided by parallel high performance computing.
This article therefore essentially depends on using the massively parallel and efficient
software frameworks \walberla\ \cite{Feichtinger2011105} and \pe{}~\cite{iglberger:2010:Pe,Iglberger:2009:CSRD}.
\walberla\ supports fluid-structure interaction with the LBM in a massively parallel setting~\cite{Goetz:2010:SC10,Godenschwager:2013:FHP}
by coupling it to the physics engine \pe{} for rigid body dynamics.
The elongated particles simulated in this paper
are modeled as spherocylinders of fixed radius and with different aspect ratios.

A validation of spherocylinder simulations with the \pe\ in absence of hydrodynamic interactions
was presented in Fischermeier~et~al.~\cite{Fischermeier20143156}.
The fluid-particle interaction algorithm with the coupled frameworks
has been developed and validated for charged
spherical particles in microfluid flows in~\cite{Bartuschat:2014:CP}.

The LBM was previously applied to study the sedimentation of elongated rigid particles at low to moderate Reynolds numbers.
Xia~et~al.~\cite{xia2009flow} simulated the settling of single elliptical particles in a narrow channel with a two-dimensional multi-block LBM method 
to examine the wall influence on flow patterns for different density, aspect, and blocking ratios.
The settling of a single spherocylinder in a channel was studied with a two-dimensional
lattice Boltzmann direct-forcing fictitious domain method in Nie~et~al.~\cite{NIE2012SedimCapsule} for different solid-fluid density ratios.
In~\cite{Jianzhong2003AspRatioSedimFiber}, the settling of cylindrical fibers was investigated for different aspect ratios at moderate Reynolds numbers.
The simulations were performed in three dimensions with the LB momentum--exchange method in a box moving with the particles.\\
The LB momentum--exchange method was previously applied to simulate the rotational motion of single elongated particles in shear flow.
Ku and Lin~\cite{Ku2009InertialEffectRot} simulated the rotation of single rigid cylinder-shaped particles in planar Couette flow for moderate Reynolds numbers.
The influence of the Reynolds number and the flow confinement was examined in two dimensions for a given aspect ratio. 
Mao and Alexeev~\cite{mao2014motion} presented three-dimensional simulations of the motion of single spheroidal particles in an unbounded shear flow at low to moderate Reynolds numbers.
In their work, the influence of fluid and particle inertia on the rotational motion was investigated
for different aspect ratios and initial orientations. 

The physical models
for elongated particles in
creeping flow are summarized in \Sect{sec:creeping}.
Details about the SBF and LBM are introduced in \Sect{sec:slender} and \Sect{sec:lbm}, respectively.
The simulation of a single elongated particle is validated in \Sect{sec:comparing},
and the tumbling motion of two sedimenting particles
is studied in \Sect{Sec:TumblingParticles}.
Finally, the findings are summarized in \Sect{sec:conclusions} and possible model extensions are outlined.
\section{Creeping flow with rod-like particles}\label{sec:creeping}
\subsection{Fluid dynamic equations}
\label{Fluiddynamicequations}

In this article we are concerned with the flow of small rigid particles suspended in a
viscous incompressible fluid.
The flow of the fluid can be described by the Navier-Stokes equations,
\begin{align}
\rho_f \left[\frac{\partial {\bf u}}{\partial t}+{\bf u}\cdot\nabla{\bf u}\right]&=-\nabla p
  + \mu_f\nabla^2 {\bf u} + {\bf f}_b, \label{eq:NS_mom}\\
\nabla \cdot {\bf u} &=0 \label{eq:NS_mass}.
\end{align}
Here, ${\bf u}$ and $p$ denote the velocity and the pressure of the fluid,
$\mu_f$ and $\rho_f$ denote the dynamic viscosity and density of the fluid, and
${\bf f}_b$ denotes the external body force density.

A density difference between the immersed particles and
the surrounding fluid will give rise to a motion of the particles.
For particles in incompressible fluids, the gravitational force acting on the fluid
does not need to be considered explicitly.
Instead, the effect of the resulting pressure gradient
and of the gravitational force acting on the particle
is accounted for in the force applied to the particle
\begin{equation}
{\bf F}=(\rho_p-\rho_f){\bf g}V,
\label{Eq:ForceDensDiff}
\end{equation}
where $\rho_p$ is the density of the particle, $V$ denotes its volume,
and ${\bf g}$ is the gravitational acceleration.

The influence of the particle on the fluid will arise from the 
boundary condition the particle imposes on the fluid motion.
For a rigid-body motion a no-slip condition is imposed meaning that the fluid velocity
at the boundary of the particle is the same as the particle velocity.
Boundary conditions must also be applied at the
outer boundaries of the domain of interest.
Which boundary conditions
to use, depends on the flow case. 
In cases when only the motion of the particle is of interest,
outer boundary conditions should be chosen that influence the flow field is as
little as possible, \eg{} periodic or free-slip boundary conditions (see \Sect{SubSec:Tumbling_Results} or \Sect{SubSec:SedimFiberLBMSetup}).

\subsection{Stokes equations}
\label{Stokesequations}
For small particles the velocity scale is usually small and therefore
also the particle Reynolds number which is defined as 
\begin{equation}
   Re_p=\frac{UL}{\nu_f}.
   \label{Eq:ParticleReynoldsNr}
\end{equation}

Here the particle Reynolds number is based on a typical length, $L$, related to the
size of the particle, as well as on a typical velocity $U$ of the particle, and
the kinematic viscosity of the fluid $\nu_f=\mu_f/\rho_f$.
In this paper, we use the particle diameter $2r$ (see \Fig{Fig:BodyGeometries}) 
as a typical length scale for the particle Reynolds number $Re_{p,d}$.

When $Re \ll 1$, the inertial and acceleration terms in the momentum
equation \eqref{eq:NS_mom} can be neglected, 
resulting in the Stokes equations
\begin{align}
\nabla p- \mu_f\nabla^2 {\bf u} &= {\bf f}_b, \label{eq:Stokes_mom}\\
\nabla \cdot {\bf u} &=0 \label{eq:Stokes_mass}.
\end{align}

\subsection{Single body motion at low Reynolds number}
\label{Sec:AnalytSolnMotionSB}
In some simple cases, e.g. one sedimenting sphere, cylinder or
spheroid in an unbounded Stokes flow, analytical solutions can be found. 
These solutions are described below.
More complex problems like interacting particles or particles in bounded fluid
domains must be solved using computational models such as LBM or SBF.

In the regime of low Reynolds numbers, a linear relation exists between the force $\Fb$ or the torque ${\bf M}$ 
applied to a particle and the resulting translational velocity ${\bf U}$ or angular velocity ${\bf \omega}$, respectively.
For a slender body with rotational symmetry around its major axis,
given by a unit vector ${\bf t}$, the corresponding equations are~\cite{Doi1986Theory} 
\begin{align}
   \label{Eq:translationalAngularVelocitySB}
   {\bf U} &= \mathbf{\Xi}^{-1}\Fb        \qquad\quad\; \text{or} &     {\bf \omega} &= {\bf M}/\gamma_r, 
\end{align}

considering only torques applied perpendicular to ${\tb}$. Here, $\gamma_r$ is the rotational friction coefficient and $\mathbf{\Xi}$ the translational friction tensor. 
The latter depends on the translational friction coefficients $\gamma^{||}_\text{t}$ and $\gamma^{\bot}_\text{t}$ 
for motion parallel and perpendicular to the symmetry axis ${\bf t}$ of the particle as~\cite{Doi1986Theory}
\begin{equation}
 \mathbf{\Xi}=\gamma^{||}_\text{t}\tb\tb^T
+\gamma^{\bot}_\text{t}({\bf I}-\tb\tb^T)\;,
\end{equation}
with identity matrix ${\bf I}$ and dyadic product $\tb\tb^T$. % of $\tb$.
For forces applied parallel (lengthwise motion) or perpendicular (sidewise motion) to $\tb$, 
the expression for ${\bf U}$ in \Eqn{Eq:translationalAngularVelocitySB} simplifies to
\begin{align}
   {\bf U^{||}}   &= \Fb/\gamma^{||}_\text{t} \qquad\quad\; \text{or} &
   {\bf U^{\bot}} &= \Fb/\gamma^{\bot}_\text{t}, 
   \label{Eq:TranslVelocitySB}
\end{align}
respectively. Thus, the velocity and angular velocity of a particle under external forces and torques can be obtained 
provided the friction coefficients are known. However, these friction coefficients depend on the geometry of the particle and are known analytically 
only in a few cases described below.
\begin{figure}[h!t]
   \centering
   \subfigure[Spherocylinder and coordinate system]{
\hspace{-1.3cm}
      \includegraphics[bb=0 0 174.08 70.95]{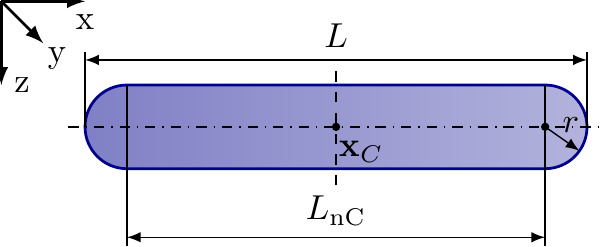} % bounding box required for arxiv, used /BBox [0 0 100 100] values in pdf file
      \label{Fig:Capsule}    
   }
   \subfigure[Cylinder]{
      \includegraphics[bb=0 0 167.94 49.99]{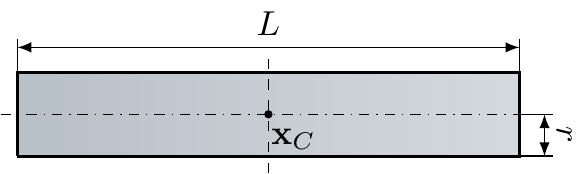}
      \label{Fig:Cylinder}
   }
   \subfigure[Spheroid]{
      \includegraphics[bb=0 0 167.94 49.99]{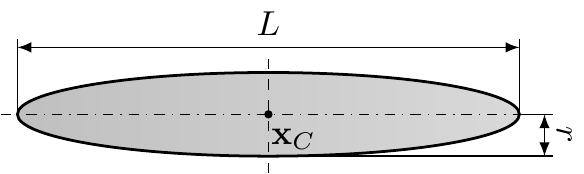}
      \label{Fig:Spheroid}
   }
   \caption{Slender body geometries and used coordinate system.} 
   \label{Fig:BodyGeometries}
\end{figure}

Cox~\cite{Cox70} derived analytical formulas for the force acting on a long, slender body at rest in Stokes flow
from an expansion of the velocity field in terms of the parameter $\eeps=r/L$. Here, $r$ is the radius of a particle and $L$ its length (see~\Fig{Fig:BodyGeometries}).
In case of a cylinder or spheroid, the translational motion can be described by the friction coefficients
\begin{align}
\gamma^{||}_\text{t} &=2\frac{\pi\mu_f L}{\ln(1/\eeps)+C_1} \quad \text{and}  & 
\gamma^{\bot}_\text{t} &=4\frac{\pi\mu_f L}{\ln(1/\eeps)+C_2}, \label{Eq:CapsTheorVelLengthwSidewMotionCox}
\end{align}
where the constants $C_1$ and $C_2$ depend on the shape of the particles.
For a circular cylinder $C_1=-3/2+\ln 2$ and $C_2=-1/2+\ln 2$ holds, for a spheroid  $C_1=-1/2$ and $C_2=+1/2$.
The above formulas by Cox are valid for large aspect ratios, due to errors in the order of $O\left( \ln\left(\eeps \right)^{-3} \right)$~\cite{Cox70}.

In their work on diffusion of cylinders~\cite{tirado1984comparison}, Tirado et al.\ 
give the diffusion coefficients $D$ for rotational diffusion, as well as diffusion parallel and perpendicular to the cylinder axis.
These diffusion coefficients are at given temperature $T$ linked to the friction coefficients $\gamma$ 
via the Einstein-Smoluchowski relation $\gamma=k_B T/ D$ with $k_B$ as Boltzmann's constant.
Using this relation, the analytical derivation in \cite{tirado1984comparison} leads to the translational frictional coefficients for lengthwise and sidewise motion, respectively
\begin{align}
   \gamma^{||}_\text{t} &= 2 \frac{\pi \mu_f L }{\ln{a} + \upsilon^{||} }  \qquad \text{and}  &
   \gamma^{\bot}_\text{t} &= 4 \frac{\pi \mu_f L }{\ln{a} + \upsilon^{\bot} },
   \label{Eq:CapsTheorVelLengthwSidesMotionTirado}
\end{align}
with $a=\frac{L}{2r} =\frac {1}{2 \eeps}$.
For rotational motion perpendicular to the cylinder axis, according to Tirado
it holds that
\begin{equation}
  \gamma_\text{r} = \frac{1}{3} \frac{\pi \mu_f L^3 }{ \ln{a} + \delta^{\bot} }.
\label{Eq:CapsTheorRotatMotionTirado}
\end{equation}
The results of Tirado et al.\ closely resemble those of Cox, however, the shape-dependent corrections $\upsilon$ and $\delta$ are functions of $a$.
In~\cite{tirado1984comparison}, Tirado et al.\ presented the following relations as simple quadratic fits in $\frac{1}{a}$ 
to the numerical values obtained in previous work~\cite{tirado1979translational,tirado1980rotational}
\begin{align}
  \upsilon^{||}   &= -0.207 +0.980 /  a -0.133 / a^2, \label{Eq:EndeffLengthwMotionTirado} \\
  \upsilon^{\bot} &=  0.839  +0.185 / a +0.233 / a^2, \label{Eq:EndeffCorrSidewMotionTirado}\\
  \delta^{\bot}   &= -0.662  +0.917 / a -0.050 / a^2. \label{Eq:EndeffCorrRotatMotionTirado}
\end{align}
With these corrections, the Tirado results are valid for aspect ratios in the range of $4 < 1/\eeps < 60$~\cite{tirado1984comparison}.

For the friction coefficients of the spherocylinders used as model particles in our LBM simulation, no analytical formulas are known. 
However, the above relations for cylinders give a good approximation for what to expect in case of spherocylinders (see \Sect{sec:comparing}).
In this article, we will use the more recent results of Tirado~et~al.\ as reference.%
\section{The slender body formulation}\label{sec:slender}
Stokes equations can be reformulated in terms of a boundary integral
equation.  In this framework the fluid velocity due to the motion of a
single particle can be computed by solving an integral equation stated
solely over the particle surface.

Consider a straight, rigid and slender particle of length $L$ and
radius $r$. If the particle has a large aspect ratio, \emph{i.e.}\
$L\gg r$, it can be referred to as a \textit{slender body}. For
slender bodies, a slender body approximation can be used.
 
\subsection{Non-local slender body approximation}
The slender body approximation is an asymptotic model derived from an
integral representation of Stokes equations.  The model relates the
velocity of the slender body's surface to forces that are consistent
with that motion and are exerted along its {\it centerline}. In the
derivation, higher-order terms in the slenderness parameter
$\eeps=r/L$
have been neglected and the accuracy of the final equation for the
velocity of the fiber center-line is of order $O(\eeps^2 \ln \eeps)$
For several interacting fibers, the accuracy is of $O(\eeps)$.  For
details on the derivation, see the work of Batchelor \cite{Ba70},
Keller and Rubinow \cite{KR76}, Johnson \cite{Jo80}, and G\"{o}tz
\cite{Go00}.

\subsubsection{Fiber velocities and force distribution}
\label{Fibervelocitiesandforcedistribution}
Assume that we have a system of $M$ fibers. Let the
center-line of each fiber be parameterized by $s\in [-l,l]$ where $l$
is the half length of the fiber. For fiber $\fno$ the coordinates of
the center-line is given by $\xb_{\fno}(t)=\xca(t)+s\pa(t)$ where
$\xca$ is the center point and $\pa$ the unit tangent vector of the
fiber and $\fno=1,2, \dots, M$. 

Assuming that the fluid exerts a force per unit length $\fb_{\fno}$
on fiber $\fno$, the slender body approximation for the velocity of the center-line of
fiber $\fno$ is given by
\begin{align}
8 \pi \mu_f (\xcadot+s\padot)= &\left[d\,(\Ib+\pa \pa^T)+2\,(\Ib-\pa \pa^T)\right]\fb_{\fno}(s) \qquad \label{eqn:sev_fils} \\
                               &+(\Ib+\pa \pa^T) \Kbarl{\fb_{\fno}}(s) + \Ucontrfno(s). \nonumber
\end{align}
Here  $d$ is a geometry parameter
\begin{equation}
d=-\ln(\varepsilon^2 e),
\label{Eq:SBFGeomParam}
\end{equation}
and $\Kbarl{\fb}(s)$  is an  integral operator given by
\begin{equation}
\Kbarl{\fb}(s)=
\int_{-l}^l
\frac{\fb(s')-\fb(s)}{|s'-s|} ds'.
\label{eqn:Kbar}
\end{equation}
The contribution to the velocity of fiber $\fno$ from the hydrodynamic
interaction of all other fibers in the system is accounted for in
$\Ucontrfno(s)$ as
\begin{equation} 
\Ucontrfno(s)=\sum_{\fnob=1}^{M}\int_{-l}^{l}
\Gb(\Rb_{\fnob \fno} (s,s')) \, \fb_{\fnob}(s')\, ds'.
\label{eqn:Ucontr_def}
\end{equation}
where  $\Rb_{\fnob \fno} (s,s')=\xca+s\pa-(\xcab+s'\pab)$ is the distance between one point
on fiber $\fno$ and one point on fiber $\fnob$.

In free-space (no outer boundary conditions) the Green's function reads
\begin{equation}
\Gb(\Rb)=\left\{\begin{array}{ll}
\Sb(\Rb) + \frac{r^2}{2}\Db(\Rb)&  \mbox{ if } \fnob \ne \fno \\
{\bf 0}, &  \mbox{ if } \fnob = \fno.
\end{array}\right.
\label{eqn:Gnonper}
\end{equation}
Here, $\Sb(\Rb) = (\Ib+\Rhat\Rhat^T)/|\Rb|$ with $\Rhat=\Rb/|\Rb|$ is the free space
Stokeslet and $\Db(\Rb)=(\Ib-3 \Rhat \Rhat^T)/|\Rb|^3$ is the dipole doublet.

The unknowns in \Eqn{eqn:sev_fils} are the translational and
the rotational velocities, $\xcadot$ and $\padot$, and the force
distribution along the fiber $\fb_{\fno}(s)$. To close the
formulation \eqref{eqn:sev_fils}, we use the additional conditions
stating that the integrated force and torque on each fiber must
balance the external forces and torques applied to the fibers
\begin{equation}
\Fb=\int_{-l}^{l} \fb_{\fno}(s) ds, \quad  \Mb=\int_{-l}^{l} s (\pa \times \fb_{\fno}(s)) ds.
\label{eqn:extra_cond} 
\end{equation}

To solve Eqns.~\eqref{eqn:sev_fils} and \eqref{eqn:extra_cond},
the force on each fiber is expanded as a sum of Legendre polynomials,
\begin{equation}
\fb_{\fno}=\frac{1}{2} \Fb + \sum_{n=1}^{N} \ab_{\fno}^n P_n(s),
\label{eqn:force_exp}
\end{equation}
where $P_n$ is a Legendre polynomial of degree $n$ and the
coefficients $\ab_{\fno}^n$ are unknown vectors with three components,
one for each direction in space. The choice of N will be a parameter
in the numerical method.
With this approach, the coefficients,
$\ab_{\fno}^n$, will be given as the solution to a dense linear system
of equations with $3MN$ unknowns. The system of equations is derived
from \Eqn{eqn:sev_fils} using the force expansion \Eqn{eqn:force_exp},
orthogonality properties of Legendre polynomials, and the fact that
the operator $\Kbar$ in \Eqn{eqn:Kbar} diagonalizes under the Legendre
polynomials, \cf{}~\cite{Go00}.  For details on the derivation, see
\cite{ToGu06}. 

Once the system of equations for the Legendre
coefficients has been solved, the force on each fiber can be
computed, and the translational and
rotational velocities for each fiber can be computed using 
\begin{equation}
  \begin{array}{r@{\hspace{0.8ex}} l}
         \xcadot=&\frac{1}{8 \pi \mu_f L} \left[d(\Ib+\pa \pa^T) + 2(\Ib-\pa \pa^T)\right]\Fb \\
                 &+\frac{1}{8 \pi \mu_f } \int_{-l}^{l} \Ucontrfno(s) \ ds, 
         \label{eqn:xadot}
  \end{array}
\end{equation}
\begin{equation}
  \begin{array}{r@{\hspace{0.8ex}} l}
      \padot=  &\frac{3d}{2 \pi \mu_f L^3}(\Mb\times \pa) \\
               &+\frac{3}{2\pi \mu_f L^3}({\bf I}-\pa \pa^T)\int_{-l}^{l}s \Ucontrfno(s) \ ds.
      \label{eqn:padot} 
  \end{array}
\end{equation}
By integrating Eqns.~\eqref{eqn:xadot} and \eqref{eqn:padot} in time, 
the position and orientations of the fibers can be updated.

\subsubsection{Numerical algorithm}
\label{sec:numtreat}

The numerical algorithm developed to solve this problem is presented
in detail in \cite{ToGu06} where also the accuracy of the numerical
method is carefully studied. Here we will only give a short summary
of the numerical algorithm.

In the numerical treatment of this problem, integrals of the form
\begin{equation}
\int_{-1}^{1}  \left[ \int_{-1}^{1}
\Gb(\Rb(s,s')) P_k(s') ds' \right] P_n(s) ds
\label{eqn:theta_def}
\end{equation}
must be computed. Note that in order to use the Legendre expansion,
the equations are solved in a dimensionless form such that $-1 \leq s
\leq 1$.
For the inner integral in \Eqn{eqn:theta_def} formulas for
analytic integration have been developed \cite{ToGu06}. The outer
integral in \Eqn{eqn:theta_def} is evaluated numerically by
splitting the integration interval into $N_q$ sub intervals, using a
three-point Gauss quadrature rule on each interval.

The linear system of equations for the coefficients in the Legendre
expansion is a dense system and is solved iteratively using GMRES
which on average converges (depending on the distance between the
fibers) within four or five iterations.

To update the position of the fibers, \Eqn{eqn:xadot} and \Eqn{eqn:padot}
are discretized in time using an
explicit second-order time-stepping scheme with a fixed time step. 

\subsubsection{Extension to periodic boundary conditions}
To perform simulations in a periodic domain, we 
must work with a periodized version of the Green's function in
\Eqn{eqn:Ucontr_def}. This term will now also include the
contribution from all periodic images of the fibers.  It has no closed
analytical form, but can be thought of as a sum over an infinite
periodic array of free space Stokeslets. 

The periodic Stokeslet is evaluated through sums
in real and Fourier space, and although rapidly converging, it is more
costly to evaluate than a Green's function with a closed analytical
expression.  Therefore, parts of the periodic Stokeslet is
initially evaluated on a uniform grid with grid size $h_g$, covering
the domain $[0,L_x/2] \times [0,L_y/2] \times [0,L_z/2]$.  Due to
symmetries, trilinear interpolation can be used to obtain the values
needed for any coordinate in a periodic box of size $[-L_x/2,L_x/2] \times
[-L_y/2,L_y/2] \times [-L_z/2,L_z/2]$.

For conditions on convergence
of this sum and its practical evaluation, see \cite{ToGu06}.

The integral over the periodized Green's function is treated
numerically in the same way as the outer integral in
\Eqn{eqn:theta_def}.%
\section{Fluid-particle interaction with the lattice Boltzmann method}\label{sec:lbm}
\subsection{The lattice Boltzmann method}
\label{sec:LBM}
The lattice Boltzmann method (LBM) is a numerical scheme for simulating hydrodynamics based on kinetic theory of gases. In the LBM,
the phase space is discretized into a Cartesian lattice $\Omega_{\textup{dx}} \subset \mathbb{R}^D$ of dimension $D$ with spacing $\textup{dx}$,
and a finite set of $Q$ discrete velocities ${\bf c}_q \in \mathbb{R}^D, q \in \{1,\ldots,Q\}$.
Associated with each ${\bf c}_q$ is a particle distribution function (PDF) $f_q: \Omega_{\textup{dx}} \times T_{\textup{dt}} \mapsto \mathbb{R}$ 
that represents the probability of an ensemble of molecules located at the lattice site ${\bf x}_i \in \Omega_{\textup{dx}}$ to move at that velocity.
These velocities are chosen such that within a time increment $\textup{dt} = t_{n+1} - t_n$
with discrete time $T_{\textup{dt}} = \{ t_n: n = 0,1,2,\ldots \} \subset \mathbb{R}^+_0$,
PDFs can move to neighboring lattice sites, or rest at a site.

The generalized discrete lattice Boltzmann equation~\cite{higuera1989lattice,dHum1992} with collision matrix $\mathbf{S}$
 \begin{equation}
\resizebox{0.885\columnwidth}{!}{$%
 f_q({\bf x}_i + {\bf c}_q \textup{dt}, t_n + \textup{dt}) - f_q({\bf x}_i,t_n) =\sum\limits_{j} S_{q j} \left( f_j - f_j^{\text{eq}} \right)
$}
   \label{Eq:discrLBE}
 \end{equation}
 is an approximation of the Boltzmann equation in discrete phase space
 that can be derived from a forward-difference discretization in time
 and a spatial upwind
 discretization~\cite{Sterling1996_StabilLBM,wolf2000lattice}. This
 equation describes the advection of PDFs between adjacent lattice
 sites and subsequent collisions that are represented by the collision
 operator in the right hand side. The equilibrium distribution
 function, $f_q^\text{eq}$, as proposed by He and
 Luo~\cite{HeLuo:97}, is given by
\begin{equation}
   \begin{array}{r@{\hspace{0.2ex}} l}
      f_q^\text{eq}(\rho_f,{\bf u}) = w_q \left[\vphantom{\frac{1}{x_X^X} {\bf x}^X}\right.  \rho_f + & \rho_0 \left( \frac{1}{c_s^2}({\bf c}_q^T {\bf u}) + \right. \\
                                                                                                      & \left.\left. \frac{1}{2c_s^4}({\bf c}_q^T  {\bf u})^2 - \frac{1}{2c_s^2}({\bf u}^T {\bf u}) \right) \right]
   \end{array}
   \label{Eq:fequ_Incompr}
\end{equation}
and recovers the incompressible Navier-Stokes (momentum) equation up to an error of order
$\mathcal O(\mbox{Ma}^3)$.
Here, $\mbox{Ma} = \frac{U}{c_s} $ is the Mach number, a
measure for compressibility that depends on the characteristic velocity
$U$ of the fluid and the thermodynamic speed of sound $c_{s}$ of the
lattice model. The equilibrium distribution function depends on the local fluid density $\rho_f({\bf x}_i,t_n) = \rho_{0} + \delta\rho({\bf x}_i,t_n)$ with average value $\rho_{0}$ and fluctuation $\delta\rho$, and up to quadratic order on the local velocity ${\bf u}$.
These macroscopic quantities can be computed as moments of $f_q$ as
\begin{equation}
  \begin{array}{l c}
      \rho_f ({\bf x}_i, t) = \sum\limits_{q} f_q({\bf x}_i, t), \\
      {\bf u}({\bf x}_i, t) = \frac{1}{\rho_{0}} \sum\limits_{q} {\bf c}_{q} f_q({\bf x}_i, t),
  \end{array}
\end{equation}
and the pressure $p$ is given by the equation of state for an ideal
gas, $p({\bf x}_i,t)=c_s^2\rho_f({\bf x}_i,t)$.

We use the D3Q19 model of~\cite{1992:QianBGK} with ${c = \textup{dx} / \textup{dt}}$,
\begin{equation}
   c_{s}= c/\sqrt{3}
   \label{Eq:SpeedOfSound}
\end{equation}
and the following weights $w_q$ for the different directions:
${w_1 = 1/3}$, ${w_{2, \ldots, 7} = 1/18}$, and ${w_{8,\ldots, 19} = 1/36}$.

We employ the
stable and accurate two-relaxation-time (TRT) collision operator by
Ginzburg~\cite{ginzburg2004lattice,ginzburg2008two},
\begin{equation}
  \sum\limits_{j} \mathbf{S}_{q j} \left( f_j - f_j^{\text{eq}} \right) =
    \lambda_e \left( f^e_q - f_q^{\text{eq},e} \right) +
    \lambda_o \left( f^o_q - f_q^{\text{eq},o} \right). 
  \label{Eq:TRTOp}
\end{equation}
with two relaxation parameters, $\lambda_e = -\tau^{-1}$
for even- and $\lambda_o$ for odd-order non-conserved moments. Here,
$\lambda_o$ is a free parameter, and the dimensionless relaxation time
$\tau$ (or collision frequency $\omega=\tau^{-1}$) is related to the
kinematic viscosity of the fluid by
\begin{equation}
\nu = \left(\tau - \frac{1}{2}\right) c_{s}^2 \textup{dt}.
\label{Eq:kinVisc}
\end{equation}
With this definition, the LBM is second order accurate in space and
time~\cite{Sterling1996_StabilLBM}.

For the TRT operator, the PDFs are decomposed as \linebreak[4] ${f_q = f^e_q + f^o_q}$ into even and odd components
\begin{equation}
  \begin{array}{l c r}
  f^e_q = \frac{1}{2} ( f_q + f_{\bar{q}} ) &\text{ and }& f^{\text{eq},e}_q = \frac{1}{2} ( f^\text{eq}_q + f^\text{eq}_{\bar{q}} ), \\[0.75ex]
  f^o_q = \frac{1}{2} ( f_q - f_{\bar{q}} ) &\text{ and }& f^{\text{eq},o}_q = \frac{1}{2} ( f^\text{eq}_q - f^\text{eq}_{\bar{q}} ),
  \end{array}
\end{equation}
with opposite velocities ${\bf c}_{\bar{q}} := - {\bf c}_q$.
The local equilibrium distribution function for the incompressible LBM
according to \cite{HeLuo:97} is then given for each lattice site by
\begin{equation}
  \begin{array}{l c}
      f^{\text{eq},e}_q = w_q \left( \rho_f - \frac{\rho_{0}}{2c_{s}^2}({{\bf u}}^T {\bf u}) + \frac{\rho_{0}}{2c_{s}^4}({{\bf c}_q}^T {\bf u})^2 \right), \\[0.75ex]
      f^{\text{eq},o}_q = w_q \frac{\rho_{0}}{c_{s}^2}({\bf c}_q^T {\bf u}).
  \end{array}
  \label{Eq:EqPDF_TRT}
\end{equation}

In each time step $t_n \in T_{\textup{dt}}$ the LBM performs a \emph{collide step} and a \emph{stream step}
\begin{equation}
  \hspace{-0.34cm}
  \begin{array}{r@{\hspace{0.55ex}}l}
  \tilde{f}_q({\bf x}_i,t_n) = f_q({\bf x}_i,t_n)  &+ \lambda_e [ f^e_q({\bf x}_i,t_n) - f_q^{\text{eq},e}({\bf x}_i,t_n) ] \\ 
                                                   &+ \lambda_o [ f^o_q({\bf x}_i,t_n) - f_q^{\text{eq},o}({\bf x}_i,t_n) ]
  \label{Eq:LBMCollide}
  \end{array}
\end{equation}
\begin{equation}
f_q({\bf x}_i + {\bf e}_q, t_n+\textup{dt}) = \tilde{f}_q({\bf x}_i,t_n),
  \label{Eq:LBMStream}
\end{equation}
where $\tilde{f}_q$ denotes the post-collision state.

% Boundary conditions and TRT parameters
Boundary conditions are treated in the stream step by modifiying the post-collision states of PDFs of fluid lattice sites ${\bf x}_{F}$ next to a boundary.
The adaption to the boundary condition is performed for the PDFs associated with directions ${\bf e}_q$, in which the neighboring cell at ${\bf x}_n = {\bf x}_{F} + {\bf e}_q$ lies on the boundary.
For the simulations presented in this article, we apply no-slip and free-slip conditions at the domain boundary.
These boundary conditions originate from lattice gas bounce-back conditions and specular reflection conditions, respectively~\cite{Cornubert1991241}.
Moreover, periodic boundary conditions are used that cyclically extend the domain in a dimension.

The no-slip boundary condition enforces zero velocity at the interface of two lattice cells by reverting the PDFs of the relevant directions as
\begin{equation}
   f_{\bar{q}}( {\bf x}_{F}, t_n + \textup{dt}) = \tilde{f}_{q}({\bf x}_{F},t_n).
   \label{Eq:BBBC}
\end{equation}
For the BGK model, the effective wall locations depend on $\tau$.
The free TRT parameter $\lambda_o$ allows to fix walls aligned with the lattice dimensions half-way between two lattice sites for $\lambda_o = - 8  (2-\omega)/(8-\omega)$~\cite{ginzburg2008study}.
This parameter is used for all simulations performed in this article.
The free-slip boundary condition enforces zero velocity in normal direction at the boundary, while retaining the tangential velocity components by reflecting the PDFs as
\begin{equation}
   f_{\text{refl(}q\text{)}}\left( {\bf x}_{F} + {\bf e}_q - {\bf n} \left( {\bf n}^T {\bf e}_q \right), t_n + \textup{dt} \right) = \tilde{f}_{q}({\bf x}_{F},t_n),
   \label{Eq:FreeSlipBC}
\end{equation}
where ${\bf n}$ denotes a normalized wall surface normal vector.
The post-reflection direction associated with $f_{\text{refl(}q\text{)}}$ can be computed as ${\bf e}_{\text{refl(}q\text{)}} = {\bf e}_q - 2 \left( {\bf n}^T {\bf e}_q \right) {\bf n}$. % This corresponds to a Householder transformation.

Periodic boundary conditions are realized by streaming the PDFs of the relevant directions to
fluid lattice sites located next to the boundary at the other side of the periodic domain.

In a domain with periodicity in all coordinate directions,
and when a constant force is applied to the embedded particles at each time step
(see \Sect{Sec:TumblingParticles}),
the system must be stabilized to prevent it from accelerating infinitely.
To keep the net momentum in the system constant, we apply a \emph{momentum stabilization technique}:
The average velocity in the whole domain is computed and then subtracted from the macroscopic velocity when computing the equilibrium distribution function.
Thus, the LBM performs relaxation towards a state with zero net momentum, while preserving all other properties.

\subsection{The momentum exchange approach}
\label{sec:MEA}
The hydrodyamic interactions of the particles via the fluid with the LBM are modeled 
by means of the momentum exchange approach,
as introduced by Ladd~\cite{Ladd_1993_PartSuspPt1,Ladd_1994_PartSuspPt2}.
This exploits the mesoscopic origin of the LBM
to compute the momentum exchange between the fluid and the suspended particles directly from PDFs adjacent to the particle boundary.
While the original method % by Ladd treats
represents particles as fluid-filled shells, we use the more stable %and accurate 
variant with solid particles according to Nguyen and Ladd~\cite{nguyen2002lubrication}.

The solid, rigid objects are mapped onto the lattice such that 
the lattice sites are divided into a set $b$ of moving obstacle cells 
whose center is overlapped by particles
and a disjoint set $F$ of fluid cells.
Thus, particles are represented as obstacle sites with a staircase approximation of their surface.
The obstacle cells $b$ that are adjacent to a fluid cell $F$ in any direction $q$
are denoted
as surface cells $s$.
The momentum transfer from particles to the fluid is modeled by
the velocity bounce-back boundary condition~\cite{Ladd_1993_PartSuspPt1}
\begin{equation}
   f_{\bar{q}} \left( {\bf x}_{F} , t_n + \textup{dt} \right) = \tilde{f}_q \left( {\bf x}_{F} , t_n \right) - 2 \frac{\omega_q}{c_s^2} \rho_{0} {\bf c}_{q}^T {\bf u}_{s}
   \label{Eq:LBMMovWall}
\end{equation}
that adapts the fluid velocity at a given fluid cell $F$ adjacent to the moving boundary
to the local velocity ${\bf u}_{s}$ at an obstacle surface cell $s$. % by adding momentum to the bounced-back PDF
The PDF that 
is bounced back from the obstacle is 
updated such that the usual no-slip boundary condition~\Eqn{Eq:BBBC} is
recovered in the stationary case, \ie{},
when the fluid velocity matches the boundary velocity~\cite{Ladd_1993_PartSuspPt1}.

The amount of momentum $\delta {\bf p}_{q}$ transferred from the fluid to a particle
within a time step
along a given link in direction ${\bf c}_q$,
can be computed (\cf{} \cite{Aidun:98:ParticulateSuspensions}) as % similar in \cite{Aidun:98:ParticulateSuspensions}
\begin{equation}
   \delta {\bf p}_{q} = \left[ {\bf c}_q  \tilde{f}_q \left( {\bf x}_{F}, t_n \right) - {\bf c}_{\bar{q}}  f_{\bar{q}} \left( {\bf x}_{F} , t_n + \textup{dt} \right) \right] \textup{dx}^3
   \label{Eq:TransferredMomentum}
\end{equation}
from the difference of the momentum densities associated with the incoming PDF of a fluid cell $F$,
and the PDF reflected from the particle surface in direction ${\bf c}_{\bar{q}}$.
The corresponding force acting on the particle along a given link can then be obtained
from the relation ${\bf F} = \frac{\delta {\bf p}}{\textup{dt}}$ and~\Eqn{Eq:TransferredMomentum} with $f_{\bar{q}}$ given by~\Eqn{Eq:LBMMovWall}.
Summing up the force contributions of the momenta transferred from fluid cells $F$ to neighbouring surface cells $s$ of a given particle
results in the overall hydrodynamic force on the particle as given in~\cite{nguyen2002lubrication}
\begin{equation}
   {\bf F}_{h} = \sum\limits_{s} \sum\limits_{q \in D_s } \left[ 2 \tilde{f}_q \left( {\bf x}_{F}, t_n \right) - 2 \frac{\omega_q}{c_s^2} \rho_{0} {\bf c}_{q}^T {\bf u}_{s} \right] {\bf c}_q  \frac{\textup{dx}^3}{\textup{dt}}.
   \label{Eq:LBMhydrodynForce}
\end{equation}
Here, ${\bf x}_{F} = {\bf x}_s + {\bf e}_{\bar{q}}$ and $D_s$ is the set of direction indices $q$, in which a given $s$ is accessed from adjacent $F$.
The overall torque ${\bf M}_{h}$ can be computed analogously 
to \Eqn{Eq:LBMhydrodynForce} by replacing  ${\bf c}_q$ by ${\bf c}_q \times \left( {\bf x}_{s} - {\bf x}_{C} \right)$,
with the particle's center of mass ${\bf x}_{C}$.

For simplicity, we use the mean density $\rho_{0}$ in~\Eqn{Eq:LBMMovWall} and consequently~\Eqn{Eq:LBMhydrodynForce} instead of the fluid density $\rho_f$ in the neighbouring fluid cell.
This is a good approximation for incompressible LBM in absence of large pressure gradients.
Moreover, as analysed in~\cite{nguyen2002lubrication},
even large deviations from $\rho_{0}$ would have negligible
effect on the accuracy of the hydrodynamic force computation.

The solid particles lead to fluid cells appearing and disappearing due to particle movement.
Lattice sites that are uncovered by a particle that is moving away 
are re-filled by setting the PDF at this site to the equilibrium distribution according to~\Eqn{Eq:EqPDF_TRT}.
We use the mean density, together with the particle surface velocity at that cell from the previous time step
to compute $f^{eq} \left( \rho_{0}, {\bf u}_{s} ( {\bf x}_s(t_{n} - \textup{dt}) \right)$.

\subsection{Coupling the LBM to rigid body dynamics}
For parallel fluid-particle interaction simulations,
the previously described methods are implemented in the 
LBM-based flow solver \walberla{} that is coupled to the physics engine \pe{}.
Both software frameworks are designed for massively parallel simulations,
using a domain partitioning approach for distributed memory parallelization with MPI.
The coupling strategy and the implementation of the fluid-particle interaction algorithm are described in~\cite{Goetz:2010:ParComp,feichtinger:2012:Diss}
and are only outlined below. % Bartuschat:2014:CP

\Walberla{}~\cite{Donath:2008:KONWIHR07,Feichtinger2011105,Bartuschat:2014:CP} 
is a parallel software framework for simulating fluid flow
that employs the LBM.
For the distributed memory parallelization, the simulation domain is decomposed into a cartesian grid of equally sized blocks that are assigned to the different MPI processes.
On each process, data from adjacent lattice sites on a neighboring process is accessible via ghost layers.
\Walberla{} performs in each time step the streaming (\Eqn{Eq:LBMStream}) and collision (\Eqn{Eq:LBMCollide}) of the LBM
that are fused to a performance-optimized stream-collide step.
Incorporated in this step is the treatment of the boundary conditions applied at the domain boundary,
together with the velocity bounce-back conditions (\Eqn{Eq:LBMMovWall}) at the particle surface
that model the momentum transfer to the fluid based on the local particle velocities % ${\bf u}_{s}$.

The \pe{}~\cite{iglberger:2010:Pe,Iglberger:2009:CSRD} is a framework for large-scale parallel rigid multi-body dynamics simulations.
The rigid objects are geometrically fully resolved, and their translational and rotational motion is computed including frictional collisions of individual particles.
Of the algorithms for multi-contact problems available in the \pe{}, we employ the parallel fast frictional dynamics (FFD) algorithm~\cite{iglberger:2010:Pe}
that is based on Kaufman et al.~\cite{Kaufman_Fast_article:2005}. 
For the parallelization, the domain is partitioned exactly as for \Walberla{}.
Each \pe{} process handles the particles whose centers of mass are located in the associated subdomain.
For the collision handling, each process additionally 
stores shadow copies of intersecting particles~\cite{preclik2015ultrascale}.
A detailed description of the parallel FFD's time-stepping procedure including MPI communication is provided in~\cite{Goetz:2010:ParComp,Fischermeier20143156}.

The momentum transfer to the particles is modeled by computing the contributions to the hydrodynamic force
at the particle surface in \Walberla{} as \Eqn{Eq:LBMhydrodynForce},
from which the \pe{} aggregates the total force acting on the center of mass and the corresponding torque.
The new positions and orientations of the particles are computed in the subsequent \pe{} step, together with their translational and angular velocities.
These velocities affect the fluid motion in the next time step, which in turn influences the particles.
This interaction modelling corresponds to a two-way coupling.

\subsection{Parallel high performance computing for the LBM}\label{sec:parallel}
The explicit time discretization of the LBM restricts the time increment,
and thus often many time steps are required to simulate physically relevant phenomena.
Due to its strictly local memory access pattern in the steam-collide step that involves only adjacent sites,
the LBM allows for highly parallel simulations with excellent scalability.
The parallel scalability of the fluid-particle interaction algorithm implemented in \walberla\
was presented in~\cite{Goetz:2010:SC10} on up to \num{294912} parallel processes.

The LBM simulations for this article were performed 
on the high performance clusters LiMa\footnote{\url{www.rrze.fau.de/dienste/arbeiten-rechnen/hpc/systeme/}} of the computing center RRZE in Erlangen (Germany)
and SuperMUC\footnote{\url{www.lrz.de/services/compute/supermuc/}} of the Leibniz Supercomputing Centre LRZ in Garching (Germany).
LiMa comprises 500 compute nodes, each containing two Xeon 5650 `Westmere' hexa-core 
processors running at 2.66 GHz and with 24 GB DDR3 RAM.
SuperMUC comprises 18 thin islands with 512 compute nodes, each node containing two Xeon E5-2680 `Sandy Bridge-EP' octa-core processors that 
are running at 2.5 GHz and that have 32 GB DDR3 RAM.
The nodes of both parallel clusters are connected by a high-speed InfiniBand interconnect.

The technical data of the parallel LBM simulations on LiMa and SuperMUC for the single particle motion validation in~\Sect{sec:comparing}
are summarized in \Tab{tab:ParallRunDataSinglePart}.
These simulations include validations of the translational and rotational motion,
examinations of wall effects, and flow field visualizations.
As overview, the minimal and maximal problem sizes and the associated parallel processes are shown,
together with the minimal and maximal time step numbers and the runtimes.
\begin{table}[h!]
   \caption{ Parallel run data for single particle motion LBM simulations.
             Lists cluster, problem size ($L_x \times L_y \times L_z$), numbers of processes (\#proc.) and time steps (\#TS), and runtime (RT)
             for validations of translational (a) and rotational (b) motion, examining wall influence on
             translational (c) and rotational (d) motion, and flow field visualization (e).
               \label{tab:ParallRunDataSinglePart} }
   \centering
   \begin{tabular}{@{\hspace{0.1ex}}l | l llll@{\hspace{0.2ex}}}
   \toprule
   
         & Cluster   & \scalebox{0.95}{$ L_x \times L_y \times L_z$}          & \#proc.                & \#TS         & RT                \\
         &           & \multicolumn{1}{c}{$[\textup{dx}]$}  &                        &              & {$[\si{\hour}]$}  \\
   \cmidrule(r){1-1} \cmidrule(lr){2-2} \cmidrule(lr){3-3} \cmidrule(lr){4-4} \cmidrule(lr){5-5} \cmidrule(lr){6-6}
   \multirow{2}{*}{a}
   &  Super-         & $2560^2 \times 2688$    & $16 \times 16 \times 32$     & \num{37000}  & $\num{8.0}$     \\
   &  MUC            &                         &                              & \num{88800}  & $\num{19}$    \\
   \cmidrule(r){1-1} \cmidrule(lr){2-2} \cmidrule(lr){3-3} \cmidrule(lr){4-4} \cmidrule(lr){5-5} \cmidrule(lr){6-6}
   \multirow{2}{*}{b}
   & LiMa     & $816 ^3$                & $8 \times 8 \times 12$       & \num{40000}  & $\num{2.6}$     \\
   &         &                         &                              & \num{140000} & $\num{9.4}$     \\
   \cmidrule(r){1-1} \cmidrule(lr){2-2} \cmidrule(lr){3-3} \cmidrule(lr){4-4} \cmidrule(lr){5-5} \cmidrule(lr){6-6}
   \multirow{5}{*}{c}
   & LiMa     & $160 ^2 \times 1200$    & $4 \times 4 \times 12$       & \num{61540}  & $\num{0.9}$     \\
   &          & $640 ^2 \times 1200$    & $8 \times 8 \times 12$       &              & $\num{4.0}$     \\
   & Super-  & $832 ^2 \times 1200$    &                              &              & $\num{6.1}$     \\
   & MUC      & $1920^2 \times 2240$    & $16 \times 16 \times 32$     & \num{86156}  & $\num{8.5}$     \\
   &          & $2560^2 \times 2688$    &                              & \num{70400}  & $\num{15}$    \\
   \cmidrule(r){1-1} \cmidrule(lr){2-2} \cmidrule(lr){3-3} \cmidrule(lr){4-4} \cmidrule(lr){5-5} \cmidrule(lr){6-6}
   \multirow{2}{*}{d}
   & LiMa     & $168 ^3$                & $2 \times 3 \times 4$        & \num{80000}  & $\num{1.3}$     \\
   &          & $1296 ^3$               & $8 \times 8 \times 12$       & \num{84000}  & $\num{19}$    \\
   \cmidrule(r){1-1} \cmidrule(lr){2-2} \cmidrule(lr){3-3} \cmidrule(lr){4-4} \cmidrule(lr){5-5} \cmidrule(lr){6-6}
   \multirow{1}{*}{e}
   & LiMa     & $832 ^2 \times 1200$    & $8 \times 8 \times 12$       & \num{70400}  & $\num{7.7}$     \\
   \bottomrule
   \end{tabular}
\end{table}

The LBM simulations of the tumbling particles in \Sect{Sec:TumblingParticles}
for the domain size of $576^3$ lattice sites and \num{600000} time steps
were performed on LiMa on $8 \times 12 \times 8$ processes and took about $\SI{16}{\hour}$.
The runs for the domain size of $768^3$ sites with \num{605000} time steps were performed on SuperMUC within \SI{48}{\hour}.
These simulations with periodic boundary conditions are slower than the single particle simulations,
due to the average velocity computation for momentum stabilization that requires global MPI communication.%
\section{Validation and comparison of the different methods for single particle motion}\label{sec:comparing}
This section 
presents a systematic validation of the models, algorithms and the software used.
To this end, we 
describe our findings regarding the motion of a single particle 
under constant force and torque, respectively, and compare the results 
obtained from LBM simulations to analytical models for slender bodies in a fluid.

\subsection{Analytical formulas}
\label{SubSec:SedimFiberAnalytFormulas}
From the analytical formulas for the motion of a slender body in
free-space given in~\Sect{Sec:AnalytSolnMotionSB}, theoretical values
for the translational and angular velocities of cylinders and
ellipsoids are computed. We compare these values to our simulation
results for validation of the implementation and used methods.

For a single fiber in a free-space setting the SBF yields explicit
formulas for the translational and angular velocity of the
fiber. Under a constant force $\Fb$ or torque $\Mb$,
Eqns. \eqref{eqn:xadot} and \eqref{eqn:padot} simplify to
\begin{align}
{\bf U}&=\xdot=\frac{1}{8 \pi \mu_f L}[d(\Ib+\tb \tb^T)+2(\Ib-\tb \tb^T)]\Fb, 
\label{eq:SBF_Onefib_vel} \\
{\bf \omega}&=\tb\times\tdot=\tb \times \frac{3d}{2 \pi \mu_f L^3}(\Mb\times \tb),
\label{eq:SBF_Onefib_rot} 
\end{align}
noting that $\Ucontrfno={\bf 0}$ when there is only one fiber 
present in the system.
These expressions are valid for an ellipsoidal particle of length $L$ and
aspect ratio $1/\varepsilon$ represented by the geometry parameter $d$
(see \Eqn{Eq:SBFGeomParam}) moving in a fluid with dynamic viscosity $\mu_f$.
Considering a force either parallel or perpendicular to $\tb$, \Eqn{eq:SBF_Onefib_vel} gives
\begin{align}
{\bf U^{||}_\text{SBF}}&=\frac{2d}{8\pi\mu_fL}\Fb    \qquad \text{and}  &
{\bf U^{\bot}_\text{SBF}}&=\frac{d+2}{8 \pi\mu_fL}\Fb.  \label{eq:SB_ForceParallPerpend}
\end{align}
These are the same expressions as derived by Cox for the lengthwise
and sidewise motion of a spheroid, see
\Eqn{Eq:CapsTheorVelLengthwSidewMotionCox}.
For $\bf M$ perpendicular to
$\tb$ \Eqn{eq:SBF_Onefib_rot} simplifies to
\begin{equation}
 {\bf \omega}_\text{SBF}=\frac{3d}{2\pi\mu_fL^3}{\bf M}\;.
\label{eq:SB_Torque}
\end{equation}

The friction coefficients for the spherocylinders
that we use in the lattice Boltzmann
simulation are not known analytically.  However, as a first
approximation we can compare our data to the analytical results for
cylinders and spheroids presented in~\Sect{Sec:AnalytSolnMotionSB}.

The  results for the spherocylinders are expected to lie between the
results for cylinders of the same radius, but with the length of the
spherocylinder without the spherical end-caps $L_\text{nC}$
($L_\text{nC} = L-2r$, see~\Fig{Fig:Capsule}) and with the full
spherocylinder length $L$.

From the analytical expressions for cylinders derived by
Tirado~et~al., the terminal translational velocity for lengthwise
motion can be computed from the friction coefficients in
\Eqn{Eq:CapsTheorVelLengthwSidesMotionTirado} with shape-dependent correction
factor in \linebreak \Eqn{Eq:EndeffLengthwMotionTirado}. In case of cylinders of
spherocylinder length without the spherical end-caps, it is denoted
by $U^{||}_\text{Tir,nC}$.  For cylinders of the full spherocylinder
length, the translational velocity is denoted by $U^{||}_\text{Tir,wC}$.

Analogously, the terminal translational velocity for sidewise motion
of cylinders computed from \Eqn{Eq:CapsTheorVelLengthwSidesMotionTirado} with
shape-dependent correction factor in \Eqn{Eq:EndeffCorrSidewMotionTirado}
is denoted by $U^{\bot}_\text{Tir,nC}$ for the spherocylinder length
without end-caps, and $U^{\bot}_\text{Tir,wC}$ for the full
spherocylinder length.
The angular velocities computed from \Eqn{Eq:translationalAngularVelocitySB} and
\Eqn{Eq:CapsTheorRotatMotionTirado} with the shape-dependent correction
factor in \Eqn{Eq:EndeffCorrRotatMotionTirado} is denoted by
$\omega_\text{Tir,nC}$ and $\omega_\text{Tir,wC}$ for the
spherocylinder length without end-caps and for the full spherocylinder
length, respectively.

\subsection{Models and parameters}
\label{SubSec:SedimFiberLBMSetup}
The parameters used for the validation experiments are chosen based on
LBM requirements in terms of spatial and temporal resolution.
To sufficiently resolve the particles, the radius
of the spherocylinders is kept constant at $r=4 \, \textup{dx}$,
with spatial discretization $\textup{dx} = \SI{10e-6}{\meter}$.
The aspect ratios $1/\eeps$ are varied from $4$ to $14$,
corresponding to particle lengths $L$ of $16 \, \textup{dx}$ to $56 \, \textup{dx}$.
As a fluid, we choose water at room temperature with kinematic viscosity
$\nu_f = \SI{1e-6}{\square\meter\per\second}$ and density $\rho_{f} = \SI{1e3}{\kilo\gram\per\cubic\meter}$,
corresponding to the dynamic viscosity $\mu_{f} = \SI{e-3}{\kilo\gram\per\meter\per\second}$.
For the LBM $\tau=6$ is chosen, which results in the time increment $\textup{dt} = \SI{183e-06}{\second}$
by its relation to $\nu_f$ and $\textup{dx}$ given in~\Eqn{Eq:kinVisc}.

The spherocylinders are modeled with the density
$\rho_\text{p} =
\SI{1195}{\kilo\gram\per\cubic\meter}$. 
To all these spherocylinders
with different aspect ratios the same force $F_z=
\SI{5.128e-10}{\kilo\gram\meter\per\square\second}$ is applied for the
translational velocity validation.
This force corresponds to the gravitational force acting on a sphere
with $r=4 \, \textup{dx}$ for the given density difference between
spherocylinders and fluid (see~\Eqn{Eq:ForceDensDiff}).
The parameters are chosen such that all simulations are performed 
with small enough Reynolds number ($\textup{Re}_{p,d} \leq 0.040$)
so that we are in the Stokes regime.
The obtained particle Reynolds numbers defined in \Eqn{Eq:ParticleReynoldsNr}
for the terminal sedimentation velocity of the spherocylinders
are in the range of $\textup{Re}_{p,d} = 0.016 \ldots 0.040$
(see \Tab{tab:TerminalSedimVel_Deviation_aspRatio_2560Domain_LBM_fluidMatWater},\ref{sec:AppendixSPM}).
Higher aspect ratios than $1/\eeps = 14$ are difficult to simulate with the LBM at very low Reynolds numbers,
due to large domain sizes required to reduce wall effects (see \Sect{SubSec:SedimFiberWallInfl})
and decreasing $\textup{dt}$ required with increasing particle lengths (see Eqns.~\eqref{Eq:ParticleReynoldsNr},~\eqref{Eq:kinVisc} and \eqref{Eq:SpeedOfSound}).

For the rotational motion validation the torques in \Tab{tab:Torques_rotatValidation_LBM_fluidMatWater} 
acting in x-direction are applied, leading to a rotation of the spherocylinder axis in the x-z plane.
These torques are chosen such that the theoretical tip velocity 
$u_\text{tip} = \SI{1e-8}{\meter\per\second}$ is the same for all aspect ratios.  
This velocity is computed as $u_\text{tip} = \omega_\text{Tir,wC} \,
L/2$, \ie{} from the angular velocity according to
\Eqn{Eq:CapsTheorRotatMotionTirado} for a cylinder with the
spherocylinder length including the end-caps.  The corresponding
theoretical particle Reynolds number is $\textup{Re}_{p,d} = 0.009$. 
The Reynolds numbers obtained in the simulations in this case are
$\textup{Re}_{p,d} = 0.013 \ldots 0.010$ for $1/\eeps = 4 \ldots 14$
(see~\Tab{tab:TerminalSedimVel_Deviation_aspRatio_2560Domain_LBM_fluidMatWater}).

\begin{table}[h!t]
  \caption{Torques $M_x$ in $\SI{e-15}{[\kilo\gram \square\meter\per\square\second]}$ applied to spherocylinders of different aspect ratios $1/\eeps$ for rotational velocity validation.
           \label{tab:Torques_rotatValidation_LBM_fluidMatWater} }
      \centering
      \begin{tabular}{l llllll}
      \toprule
   $1/\eeps$              &   4       &   6       &   8       &   10      &   12      &   14     \\
      \midrule
   $M_x$                  &   12.26   &   17.86   &   24.62   &   32.38   &   41.09   &   50.68  \\
      \bottomrule  
      \end{tabular}
\end{table}

For the LBM simulations, the particles are placed in large domains to
minimize the wall influence and to make the results comparable to
free space. The translational velocity validation with the LBM is
performed in a cuboid domain with free-slip boundary conditions
applied at the walls. The size of the domain in the direction of
motion is elongated and denoted by $L_z$. In the other dimensions it
has the same size, \ie{}, $L_x = L_y$.
Initially, the spherocylinder is
placed at the center w.r.t.\ the $x$- and $y$-dimension, and at a
distance of $z=100 \, r$ from the top boundary of the domain
located at $z=0$. 
Here, the coordinate system depicted in \Fig{Fig:Capsule} is used,
and a slice through the simulation domain center along the x-z plane is shown in \Fig{fig:SedimentCaps_FlowField}.
Under the influence of $F_z$, the spherocylinder is
then moving downwards along the domain centerline in positive z-direction.

For validation of the rotational velocity,
the spherocylinder is placed
at the center of a cubic domain, and the torques $M_x$ from
\Tab{tab:Torques_rotatValidation_LBM_fluidMatWater} are applied constantly.
Since these torques act in tangential direction to the domain boundaries,
no-slip boundary conditions are applied at the walls
to prevent the fluid from accelerating infinitely.
The same setups are used for studying the influence of the wall on the terminal motion and
for the flowfield visualization. 

\subsection{Single particle motion results}
\label{SubSec:SedimFiberResults}
We validate the LBM by comparing the spherocylinder velocities
to analytical solutions according to the theories of Cox and Tirado~et~al.\
for cylinders, as well as the SBF for ellipsoids.
Moreover, the influence of the particle shape, fluid inertia, and wall effects on the particle velocities
is investigated for the different methods and theories.

For the translational velocity validations, the simulations are performed
for the domain size $\left[2560 \, \textup{dx} \right]^2 \times 2688 \, \textup{dx}$
for \num{37000} to \num{64400} time steps for the lengthwise moving particles, and for \num{56200} to \num{88800} time steps for the sidewise motion.
A high number of time steps is required until
the particles reach steady-state---the longer the particles are, 
the more time steps are needed.
For the validation of the angular velocity, 
the LBM simulations are performed for the domain size $\left[816 \, \textup{dx} \right]^3$ 
and \num{40000} to \num{140000} time steps.

\begin{figure}[h!]
\centering
%       \includestandalone[mode=buildnew]{./FIGURES/sinkVelValidations/csvFig/TerminalSedimVel_LengthwMovCaps_aspRatio_LBM_fluidMatWater}
      \includegraphics[bb=0 0 246.7 183.45]{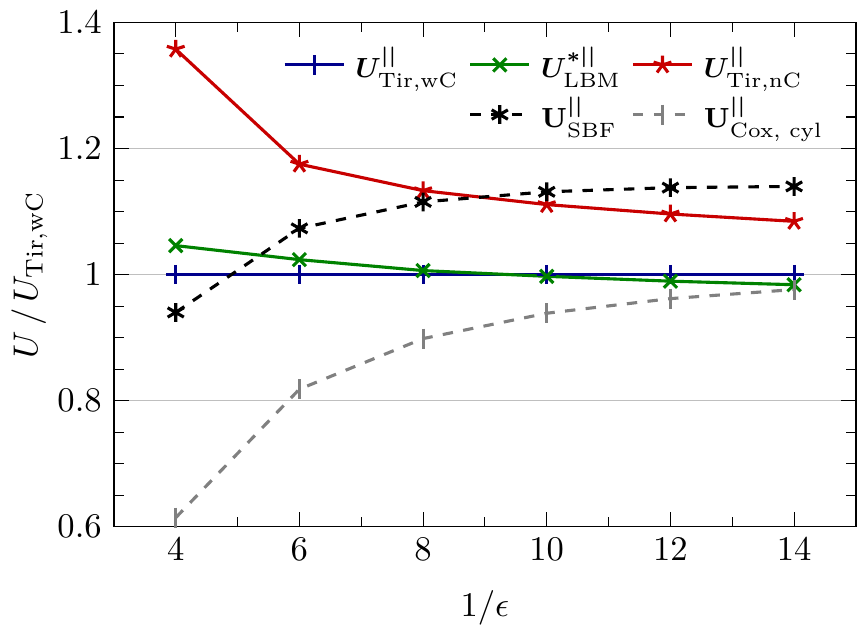}
  \caption{ Normalized sedimentation velocities $U / U_\text{Tir,wC}$ for lengthwise orientation w.r.t. direction of force $F_z= \SI{5.128e-10}{\kilo\gram\meter\per\square\second}$
            for aspect ratios $1/\eeps$ and radius $r=4 \,\textup{dx}$.
            Comparison of LBM velocities ($U^{*||}_\text{LBM}$) in $\left[2560 \, \textup{dx} \right]^2 \times 2688 \, \textup{dx}$ sized
            domain with free-slip boundaries,
            to free-space solutions by Tirado for cylinders of lengths including ($U^{||}_\text{Tir,wC}$) and excluding ($U^{||}_\text{Tir,nC}$) spherocylinder end-caps,
            and to Cox (${\bf U^{||}_\text{Cox, cyl}}$) and SBF (${\bf U^{||}_\text{SBF}}$) for cylinders and ellipsoids of full spherocylinder lengths, respectively.
           \label{fig:TerminalSedimVel_LengthwMovCaps_aspRatio_LBM_fluidMatWater} }
\end{figure}

\begin{figure}[h!]
\centering
%     \includestandalone[mode=buildnew]{./FIGURES/sinkVelValidations/csvFig/TerminalSedimVel_SideMovCaps_aspRatio_LBM_fluidMatWater}
      \includegraphics[bb=0 0 246.7 177.19]{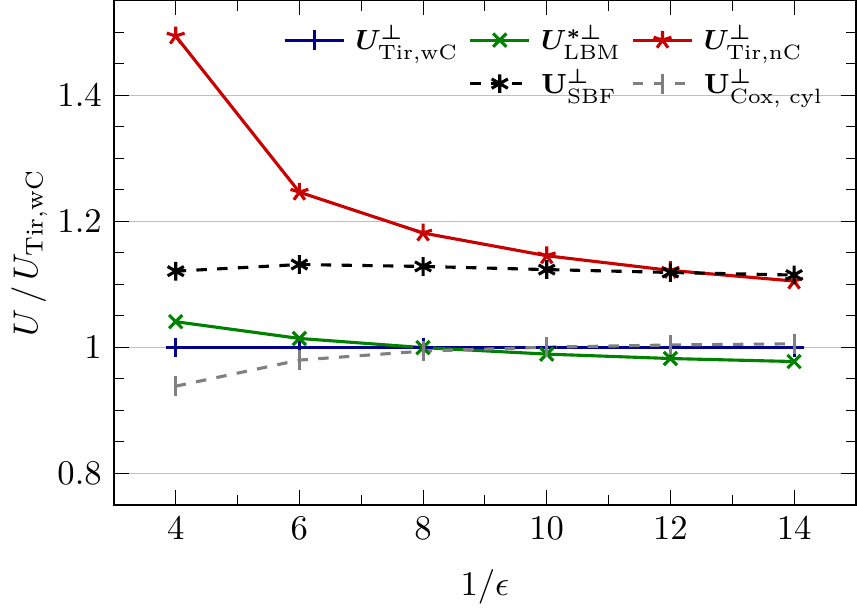}
  \caption{ Normalized sedimentation velocities $U / U_\text{Tir,wC}$
            for orientation perpendicular to $F_z$,
            $r=4 \,\textup{dx}$, and aspect ratios $1/\eeps$.
            Comparison of LBM velocities ($U^{*\bot}_\text{LBM}$) in $\left[2560 \, \textup{dx} \right]^2 \times 2688 \, \textup{dx}$
            domain with free-slip boundaries,
            to free-space solutions by Tirado ($U^{\bot}_\text{Tir,wC}$, $U^{\bot}_\text{Tir,nC}$), Cox (${\bf U^{\bot}_\text{Cox, cyl}}$),
            and SBF (${\bf U^{\bot}_\text{SBF}}$).
           \label{fig:TerminalSedimVel_SidewMovCaps_aspRatio_LBM_fluidMatWater} }
\end{figure}

\begin{figure}[h!t]
  \centering
%     \includestandalone[mode=buildnew]{./FIGURES/sinkVelValidations/csvFig/TerminalRotVel_aspRatio_LBM_fluidMatWater}
      \includegraphics[bb=0 0 246.7 183.45]{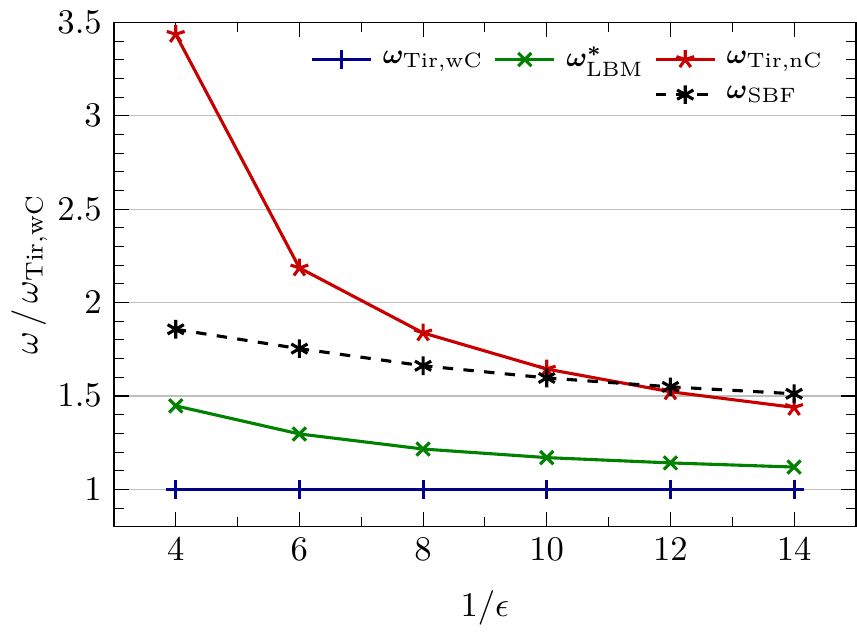}
  \caption{ Normalized angular velocities $\omega / \omega_\text{Tir,wC}$ of particles with $r=4 \, \textup{dx}$ and aspect ratios $1/\eeps$,
            resulting from torques $M_x$ in \Tab{tab:Torques_rotatValidation_LBM_fluidMatWater}.
            Comparison of LBM velocities ($\omega^{*}_\text{LBM}$) in $\left[816 \, \textup{dx} \right]^3$ domain with no-slip boundaries,
            to free-space solutions by Tirado ($\omega_\text{Tir,wC}$, $\omega_\text{Tir,nC}$)
            and the SBF (${\bf \omega}_\text{SBF}$).
           \label{fig:TerminalRotVel_aspRatio_LBM_fluidMatWater} }
\end{figure}

The sedimentation velocities for lengthwise and sidewise motion are shown in \Fig{fig:TerminalSedimVel_LengthwMovCaps_aspRatio_LBM_fluidMatWater}
and \Fig{fig:TerminalSedimVel_SidewMovCaps_aspRatio_LBM_fluidMatWater}, respectively,
for spherocylinders, cylinders, and ellipsoids with different aspect ratios.
The angular velocities are presented in \Fig{fig:TerminalRotVel_aspRatio_LBM_fluidMatWater}.
For the considered aspect ratios, the analytical expressions by Tirado~et~al.\ are the most accurate,
wheras the asymptotic expressions by Cox and the SBF are valid for very high aspect ratios (see \Sect{Sec:AnalytSolnMotionSB}).
Thus, the obtained velocities are normalized by the solutions $U_\text{Tir,wC}$ and $\omega_\text{Tir,wC}$ of Tirado~et~al.\ for cylinders of the full particle length,
in order to highlight the differences in the velocities.
% \pgfplotstableread[comment chars={\%}]{./FIGURES/sinkVelValidations/LBM_TermSedimVelSI_fluidMatWater_LengthwMovCaps_4dx_size2560.csv}\loadedtableCapsulesLengthwMov

In the figures, terminal sedimentation velocities from
the previously described LBM simulations 
for lengthwise  ($U^{*||}_\text{LBM}$), sidewise  ($U^{*\bot}_\text{LBM}$), and rotational ($\omega^{*}_\text{LBM}$)
motion are plotted, normalized by the corresponding Tirado velocities.
For comparison, the normalized velocities according to Tirado~et~al.\ for cylinders of the same total length (denoted by subscript `$\text{Tir,wC}$'), 
and for the spherocylinder length without the spherical end-caps are shown (`$\text{Tir,nC}$'), together with the Cox results (`$\text{Cox, cyl}$') for cylinders and the SBF results (`$\text{SBF}$').
% \pgfplotstableread[comment chars={\%}]{./FIGURES/sinkVelValidations/LBM_TermSedimVelSI_fluidMatWater_SidewMovCaps_4dx_size2560.csv}\loadedtableCapsulesSidewMov

\begin{figure}[h!t]
\centering
%     \includestandalone[mode=buildnew]{./FIGURES/sinkVelValidations/csvFig/TerminalSedimVel_LengthwMovCaps_largeAspRatio_LBM_fluidMatWater}
      \includegraphics[bb=0 0 255.68 183.45]{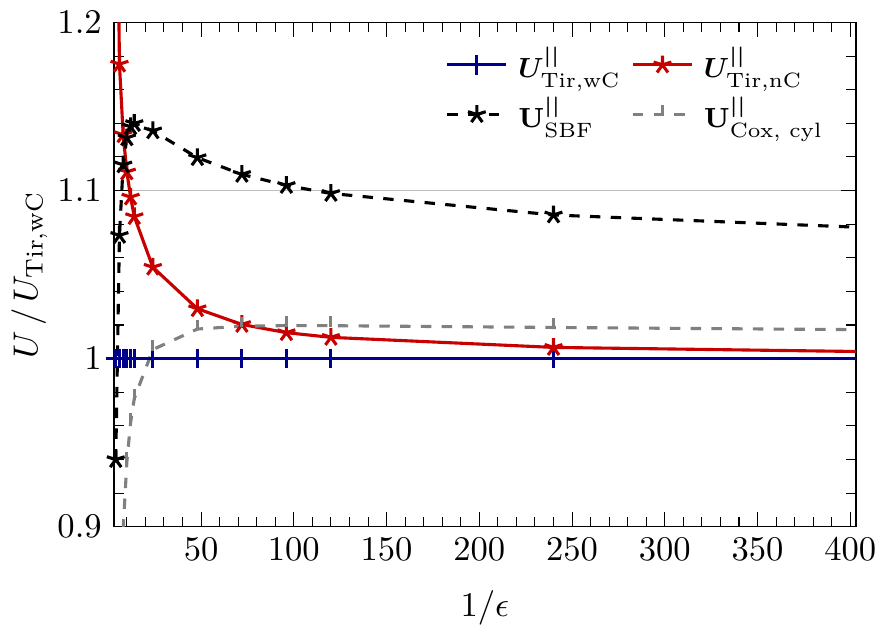}
  \caption{ Normalized sedimentation velocities $U / U_\text{Tir,wC}$ for lengthwise orientation w.r.t.\ $F_z$
            for $r=4 \,\textup{dx}$ and high aspect ratios $1/\eeps$. %  and radius $r=4 \,\textup{dx}$, $\textup{dx} = \SI{e-5}{\meter}$,
            Comparison of theories for cylinders by Tirado ($U^{||}_\text{Tir,wC}$, $U^{||}_\text{Tir,nC}$) and Cox (${\bf U^{||}_\text{Cox,cyl}}$) and 
            for ellipsoids by the SBF (${\bf U^{||}_\text{SBF}}$).
           \label{fig:TerminalSedimVel_LengthwMovCaps_largeAspRatio_LBM_fluidMatWater} }
\end{figure}

The LBM results in
\Fig{fig:TerminalSedimVel_LengthwMovCaps_aspRatio_LBM_fluidMatWater}
to \Fig{fig:TerminalRotVel_aspRatio_LBM_fluidMatWater} are mean values
of particle velocities after a sufficient number of time steps so that
steady state is reached with sufficient accuracy.
Due to obstacle mapping effects, the particle velocities fluctuate
as the effective particle volume varies with the number of overlapped particle cells (see \Sect{sec:MEA}).
The mean values $U^{*}_\text{LBM}$ and $\omega^{*}_\text{LBM}$ are
computed from the velocities 
sampled every $200$ time steps.  For the
validation of translational velocity,
the last 15\% of these
velocity values
are considered and the last 50\% for the 
validation of the angular velocity
that converges to steady state more quickly.

The fluctuations are computed from the minimum and maximum values of
the considered velocities as $\delta_\text{U} = \linebreak
({U^*_\text{max}-U^*_\text{min}})/{U^*_\text{LBM}}$ and
$\delta_\omega$, analogously.
The fluctuations for the translational motion are lower
than for the rotational motion.  For sidewise moving spherocylinders,
fluctuations of $\delta_\text{U} = 0.8\%$ occur for all aspect ratios.  For the lengthwise
moving spherocylinders, the fluctuations correspond to a value of $\delta_\text{U} = 0.8\%$ for $1/\eeps = 4$
and decrease with increasing particle length. 
For the rotational motion, fluctuations of $\delta_\omega = 8\%$ arise for $1/\eeps=4$ and 
decrease to $\delta_\omega = 2.8\%$ for the longest examined
particle.  Exact figures of the fluctuations are given in
\Tab{tab:TerminalSedimVel_Deviation_aspRatio_2560Domain_LBM_fluidMatWater}
(\ref{sec:AppendixSPM}), together with figures of the LBM results
and associated Reynolds numbers, of the analytical solutions according
to Tirado~et~al.\, and of the relative deviations
%. The deviations are computed as 
$\Delta_\text{r}U = ({U^*_\text{LBM}-U_\text{Tir}})/{U_\text{Tir}}$ and $\Delta_\text{r,}\omega$ (defined analogously)
of the LBM results from the Tirado velocities.

The terminal velocities for translational motion of spherocylinders 
in \Fig{fig:TerminalSedimVel_LengthwMovCaps_aspRatio_LBM_fluidMatWater}
and \Fig{fig:TerminalSedimVel_SidewMovCaps_aspRatio_LBM_fluidMatWater}
are close to the cylinder velocities according to Tirado for full
spherocylinder lengths. 
For small aspect ratios $1/\eeps \leq 8$ the spherocylinder velocities are slightly higher than $U_\text{Tir,wC}$
and are thus closer to those of cylinders with the length of of the spherocylinder 
without the end-caps. 
The LBM velocity for $1/\eeps =4$ is by $\Delta_\text{r}U = 4.6\%$ higher than
$U_\text{Tir,wC}$ for lengthwise motion and by 4.1\% for sidewise motion (see \Tab{tab:TerminalSedimVel_Deviation_aspRatio_2560Domain_LBM_fluidMatWater}).
For large aspect ratios, the LBM sedimentation
velocities are slightly lower than the analytical solutions
$U_\text{Tir,wC}$ for cylinders. 
The LBM velocity for the highest aspect ratio is by $\Delta_\text{r}U =1.6\%$ 
lower than $U_\text{Tir,wC}$ for lengthwise motion and
by 2.3\% for sidewise motion.\\
The lower spherocylinder velocities for $1/\eeps \geq 10$ result from wall effects
that slow down the particles compared to a free-space setting.
These effects are examined more closely in~\Sect{SubSec:SedimFiberWallInfl}.
The wall influence is higher for sidewise orientation, 
resulting in LBM results and the Tirado results $U_\text{Tir,wC}$ 
to coincide for the lower aspect ratio of $1/\eeps=8$,
compared to $1/\eeps=10$ for lengthwise orientation.\\

Inertial effects are included in the LBM simulations and result in
lower sedimentation velocities compared to Stokes flow
since the fluid resistance increases with Reynolds number.
For both, lengthwise and sidewise motion, the LBM velocities $U^{*}_\text{LBM}$ are higher than $U_\text{Tir,wC}$ only for low aspect ratios,
\ie{}, for minimal Reynolds numbers based on the particle length.
For long particles $U^{*}_\text{LBM}$ is lower than $U_\text{Tir,wC}$, and thus inertia might play an additional role in the translational motion simulations.\\
For both orientations, the velocities according to 
Cox and the SBF exhibit a similar behavior for small aspect ratios.
Both velocities increase at the same rate with increasing particle lengths for low aspect ratios and slowly decrease for higher aspect ratios.
However, while Cox' results quickly converge to values close to the Tirado velocities, the SBF velocities of ellipsoids stay higher.
For $8 \leq 1/\eeps \leq 14$, the SBF velocities are by $10-15 \%$ higher than the results by Tirado.
%
% \pgfplotstableread[comment chars={\%}]{./FIGURES/sinkVelValidations/TermSedimVelTiradoSI_fluidMatWater_LengthwMovCaps_4dx_largeAspRatio.csv}\loadedtableCylLengthwMovLrgAspRatio
This effect can be attributed to the different particle shapes, which becomes clear in
\Fig{fig:TerminalSedimVel_LengthwMovCaps_largeAspRatio_LBM_fluidMatWater} that shows the velocities according to the different theories 
for lengthwise motion at higher aspect ratios than simulated with the LBM.
Here, the SBF velocities of ellipsoids are first higher than the velocities according to Tirado 
for $1/\eeps \geq 8$ and converge only very slowly towards the results for cylinders.
For sidewise motion the differences between the theories are smaller, and the associated velocities converge faster than for the lengthwise motion,
due to a lower dependence on the particle shape.

For the rotational motion of elongated particles shown in \Fig{fig:TerminalRotVel_aspRatio_LBM_fluidMatWater}
the differences between the theories and methods are larger than for translational motion.
% \pgfplotstableread[comment chars={\%}]{./FIGURES/sinkVelValidations/LBM_CapsRotatVelSI_fluidMatWater_4dx_size816_SBF.csv}\loadedtableCapsulesRot
Also the Tirado results depend more strongly on the particle length.
As for translational motion, the velocity according to the SBF is higher than the Tirado solution for cylinders with full spherocylinder length.
The angular velocities of the spherocylinders from the LBM simulations
lie between the solutions for cylinders of a length with and without
the end-caps for all aspect ratios.
Again, the LBM velocities are
closer to the values of a cylinder of the full spherocylinder length.
However, the relative deviation from $\omega_\text{Tir,wC}$ is 
significantly higher than for the translational
motion. For $1/\eeps =4$, $\Delta_\text{r,}\omega=45\%$ and for
$1/\eeps =14$ still $\Delta_\text{r,}\omega=12\%$.
With increasing aspect ratio, the LBM velocities more closely approach the Tirado velocities with the full spherocylinder length than the SBF results do.
The angular velocities by the SBF exceed the Tirado results by almost $90\%$ for $1/\eeps=4$ and still by more than $50\%$ for $1/\eeps =14$.
The wall influence is negligible for rotational motion, as shown in~\Sect{SubSec:SedimFiberWallInfl}.
With increasing particle length, the angular LBM velocities $\omega^{*}_\text{LBM}$ decrease towards $\omega_\text{Tir,wC}$.
Thus, also for rotational motion the LBM results might be influenced by inertial effects that lead to lower velocities,
in addition to the higher hydrodynamic similarity of longer spherocylinders and cylinders of same length.

\subsection{Wall influence in LBM simulations}
\label{SubSec:SedimFiberWallInfl}
\begin{figure}[h!]
  \centering
%     \includestandalone[mode=buildnew]{./FIGURES/sinkVelValidations/csvFig/relDeviatTerminalVelCaps_vs_DomainSize_LBM_fluidMatWater}
      \includegraphics[bb=0 0 263.16 200.58]{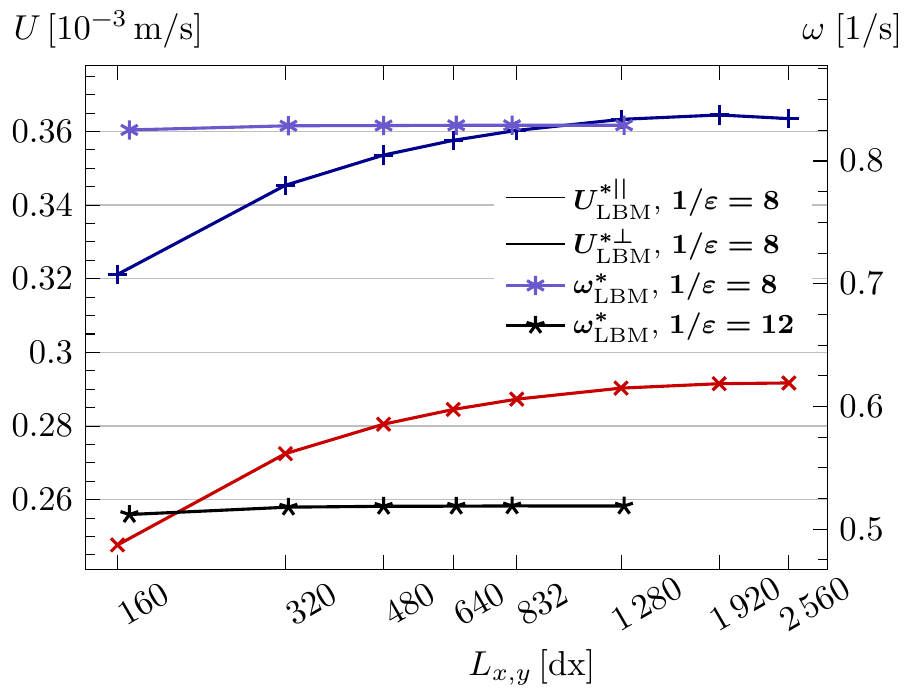}
  \caption{ Sedimentation and rotation velocities of spherocylinders with radius $r=4 \,\textup{dx}$ % , $\textup{dx} = \SI{e-5}{\meter}$
            and aspect ratio $1/\eeps =8$---and $1/\eeps =12$ for rotation---from LBM simulations in water-filled domains of different sizes ($L_{x,y}$).
            Domains for the sedimentation velocities of lengthwise ($U^{*||}_\text{LBM}$) and perpendicular ($U^{*\bot}_\text{LBM}$) oriented spherocylinders 
            (w.r.t. force $F_z$)
            are cuboids with free-slip boundaries,
            and cubes with no-slip boundaries for the angular velocities ($\omega^{*}_\text{LBM}$) resulting from torques $M_x$ in \Tab{tab:Torques_rotatValidation_LBM_fluidMatWater}.
            \label{fig:relDeviatTerminalVelCaps_vs_DomainSize_LBM_fluidMatWater}
         }
\end{figure}

The influence of the wall effect in the LBM simulations
on the settling velocity of spherocylinders
is examined for domains with quadratic cross-section of $L_x = L_y := (160, 320,\linebreak[1] 480, 640, 832) \, \textup{dx}$ and constant length $L_z = 1200 \, \textup{dx}$,
and with sizes $\left[1280 \, \textup{dx} \right]^2 \times 1600 \, \textup{dx}$, 
$\left[1920 \, \textup{dx} \right]^2 \times 2240 \, \textup{dx}$, and \linebreak[1] $\left[2560 \, \textup{dx} \right]^2 \times 2688 \, \textup{dx}$.
The simulations investigating the wall influence on rotational motion are performed
for cubic domains with edge lengths of $(168, 324, 480, 648, 816, 1296) \, \textup{dx}$.
For the aspect ratios $1/\eeps=8$ and $1/\eeps=12$, \num{80000} and \num{84000} time steps are performed, respectively.\\
The terminal translational and rotational velocities are presented
in~\Fig{fig:relDeviatTerminalVelCaps_vs_DomainSize_LBM_fluidMatWater}
on the left and right ordinate, respectively, dependent on $L_x =L_y$
plotted as `$L_{x,y}$' on the abscissa. 
The displayed velocities are again
the mean values of the particle velocities 
evaluated every $200$ time steps.
For the translational motion, the last 34\% of these velocities are
considered, and the last 15\% for the largest domain only. For the
rotational motion, the last 50\% are considered.
%
% \pgfplotstableread[comment chars={t}]{./FIGURES/sinkVelValidations/LBM_TermSedimVelSI_fluidMatWater_wC_EpsInv8_sevDomSizes.csv}\ldTabVel
% \pgfplotstableread[]{./FIGURES/sinkVelValidations/LBM_AngularVelSI_fluidMatWater_wC_EpsInv8_12_sevDomSizes.csv}\ldTabAngVel

The terminal translational velocities of the spherocylinders with 
$1/\eeps=8$ in~\Fig{fig:relDeviatTerminalVelCaps_vs_DomainSize_LBM_fluidMatWater}
increase significantly with domain size.  For the translational motion
of lengthwise oriented spherocylinders, the retarding effect of the
confined domain on the sedimentation velocity is negligible for domain
sizes of $\left[1280 \, \textup{dx} \right]^2 \times 1600 \,
\textup{dx}$ and above.  For sidewise oriented particles, only the
$\left[2560 \, \textup{dx} \right]^2 \times 2688 \, \textup{dx}$
domain is sufficiently large to assume negligible wall effects.
These findings confirm the statement in \Sect{SubSec:SedimFiberResults} that
the wall effect is measurable for $1/\eeps \geq 10$ in the 
$\left[2560 \, \textup{dx} \right]^2 \times 2688 \, \textup{dx}$ domain, and
that the corresponding velocities are under-estimated.

For the rotational motion, there is hardly any impact of the confined domains
on the angular velocities for both aspect ratios $1/\eeps=8$ and
$1/\eeps=12$.  Thus, the wall effect for the $\left[816 \, \textup{dx}
\right]^3$ domain in \Sect{SubSec:SedimFiberResults} is negligible for
all considered aspect ratios.  According to the angular velocities shown
in~\Fig{fig:relDeviatTerminalVelCaps_vs_DomainSize_LBM_fluidMatWater},
domains with edge length $480 \, \textup{dx}$ and $648 \, \textup{dx}$
are sufficient for ${1/\eeps=8}$ and ${1/\eeps=12}$, respectively.

\subsection{Flow field around sedimenting particles}
\label{SubSec:SedimFiberFlowField}
\begin{figure*}[h!tb]
  \centering
  \subfigure[Lengthwise moving spherocylinder after \num{22000} time steps.\label{fig:LengthwiseMovCaps_FlowField}]{
  \centering
  \includegraphics[bb=0 0 1600 852, trim = 44.5mm 50.5mm 215mm 39.5mm, clip, width=0.55\linewidth,keepaspectratio, angle=270]{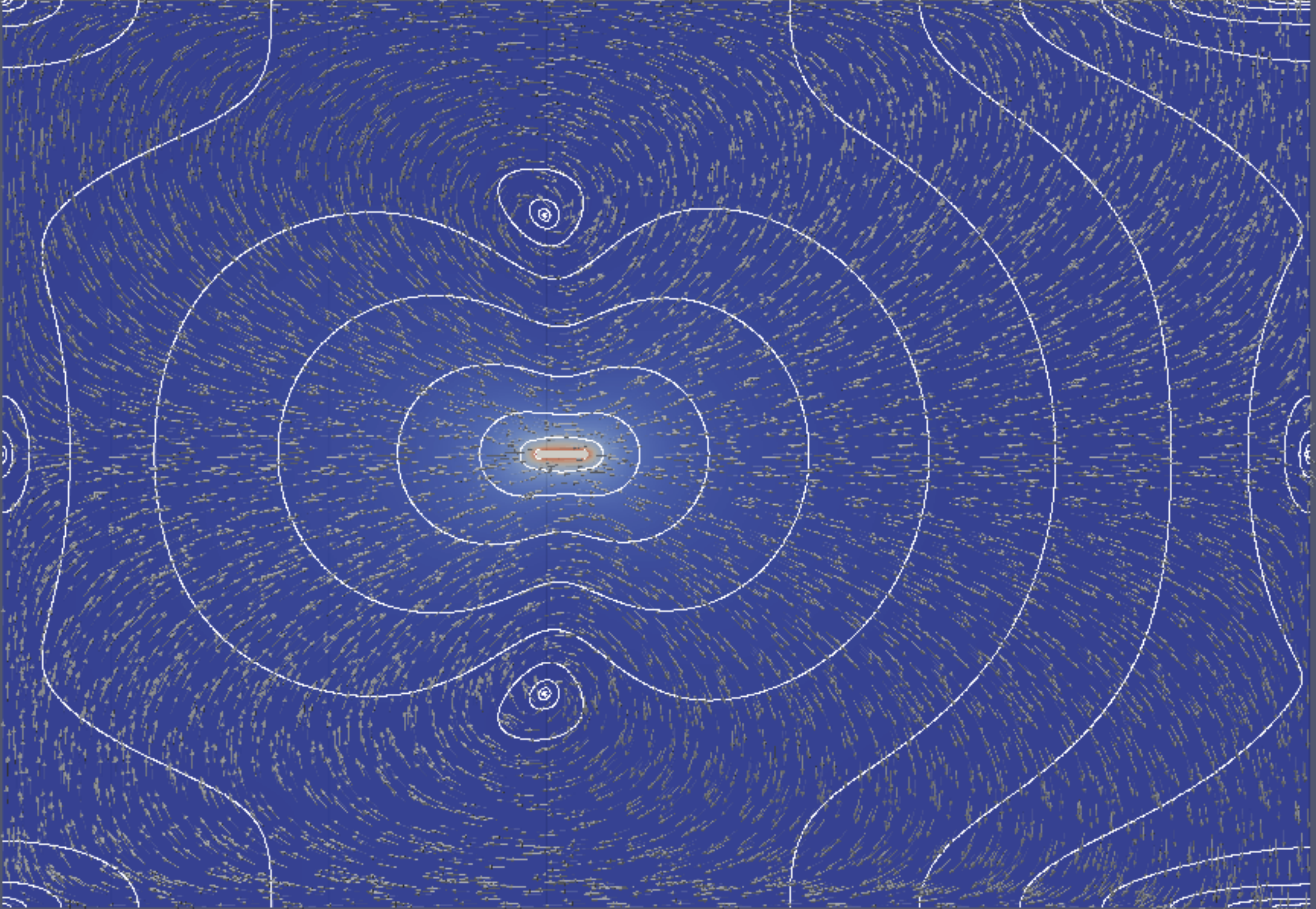} %trim = 2mm 2mm 162mm 2mm
  }\hspace{5mm}
  \subfigure[Sidewise moving spherocylinder after \num{28000} time steps. \label{fig:SidewiseMovCaps_FlowField}]{
  \centering
  \includegraphics[bb=0 0 1600 852, trim = 44.5mm 50.5mm 215mm 39.5mm, clip, width=0.55\linewidth,keepaspectratio, angle=270]{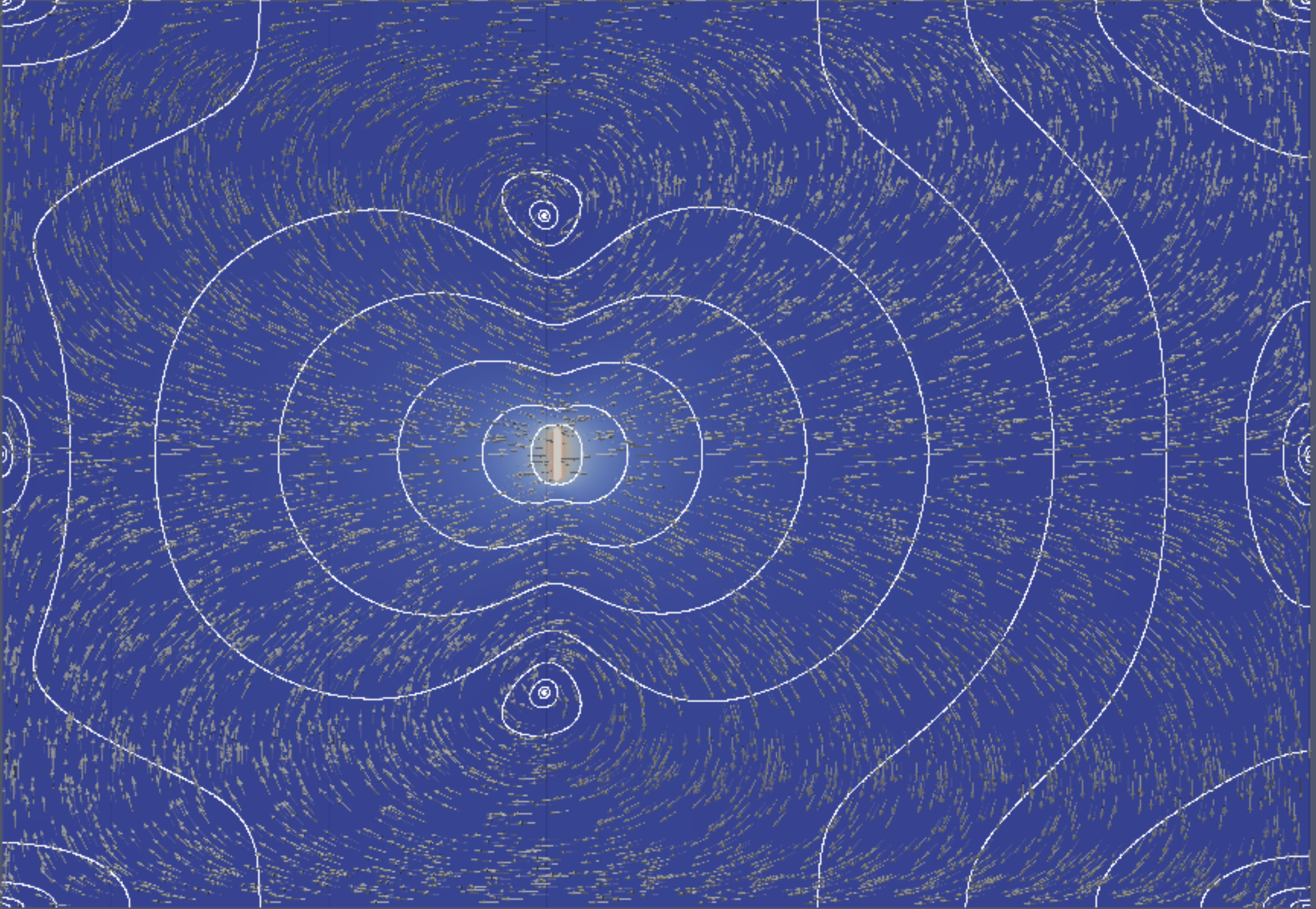}
  }
  \caption{ Flow around sedimenting spherocylinders along x-z plane through domain center.
            Flow direction is indicated by small streaks, and velocity magnitude by color in range of \SI{360e-6}{\meter\per\second} (red) to \SI{0}{\meter\per\second} (blue)
            and logarithmic contour lines.\label{fig:SedimentCaps_FlowField}}
\end{figure*}

We present the flow field around a single sedimenting spherocylinder
in a closed domain with free-slip boundary
conditions simulated with the LBM.
The simulations are performed for spherocylinders of aspect ratio $1/\eeps=12$
in a domain of size $\left[832 \, \textup{dx} \right]^2 \times 1200 \, \textup{dx}$ for \num{74800} time steps.\\
The fluid velocity around the
sedimenting spherocylinder and the staircase-approximated
spherocylinder itself depicted in \Fig{fig:SedimentCaps_FlowField} are visualized with ParaView.
The flow field around the spherocylinder is shown along the x-z plane 
through the domain center for a particle oriented lengthwise and sidewise along the direction of the
applied force $F_z$ in \Fig{fig:LengthwiseMovCaps_FlowField} and \Fig{fig:SidewiseMovCaps_FlowField},
respectively.
Since the velocity magnitude decays quickly with increasing distance from the
spherocylinders, white isosurface contour lines indicate the velocity
magnitude with logarithmic contour intervals. The flow direction is
depicted by small streaks of uniform length.
The sixteen contour lines are plotted for both orientations at velocity magnitudes 
in the range of \SI{304e-6}{\meter\per\second} to \SI{304e-12}{\meter\per\second}.

The initially quiescent fluid is dragged along by the spherocylinder
and starts moving adjacent to the particle.  Along the channel
centerline a fluid flow develops in the direction of spherocylinder motion.  
The walls in z-direction cause a counterflux and thus a
vortex. Initially the vortex is located next to the particle and
then moves to a position half-way between particle surface and
confining walls. For the lengthwise oriented spherocylinder
in~\Fig{fig:LengthwiseMovCaps_FlowField}, the contour lines have a
similar shape as the particle.
For the sidewise oriented spherocylinder
in~\Fig{fig:SidewiseMovCaps_FlowField}, the
contour lines next to the particle are circular, and they have similar
shape as for the lengthwise oriented spherocylinder in some distance.
However, the shape is more circular, \ie{}, the flow is influenced to
a higher degree also orthogonal to the movement direction.
Moreover, the sidewise moving spherocylinder influences the fluid at larger distances
than the lengthwise oriented spherocylinder,
as can be seen from the distance of equivalent contour lines to the particles.

In accordance with the results in~\Sect{SubSec:SedimFiberResults}, the
particle with sidewise orientation moves slower than for lengthwise
orientation.  The same position in movement direction is reached after
\num{22000} time steps for lengthwise orientation, and after
\num{28000} time steps for sidewise motion.

The flow field caused by the sidewise moving spherocylinder 
is nearly equal in both directions orthogonal to the movement direction.
Only very close to the spherocylinder, the fluid velocity in x-z plane shown in \Fig{fig:SidewiseMovCaps_FlowField}
is found to be slightly higher than in the orthogonal y-z plane.%
\section{Tumbling particles}
\label{Sec:TumblingParticles}
In this section we study the tumbling motion of two elongated particles in a periodic domain. 
We present simulation results for spherocylinders with the LBM, as well as SBF results for ellipsoidal fibers.
A comparison of the results from the two methods is presented for different domain sizes and aspect ratios.
For the LBM, the influence of the initial particle distance on the tumbling behavior is examined.

\subsection{Setup and parameters}
\label{SubSec:Tumbling_SetupParam}
The LBM simulations are performed with two spherocylinders of radius $r=4 \, \textup{dx}$, with ${\textup{dx} = \SI{4.98e-6}{\meter}}$.
The particle length, $L$, is varied between $40  \, \textup{dx}$ and $56  \, \textup{dx}$, corresponding to the aspect ratios $1/ \eeps = 10$ to $1/ \eeps = 14$.
Initially the particles are placed, aligned with the direction of gravity,
in a periodic box at the center w.r.t.\ y-direction and
centered around the middle of the domain in x-direction with a given center-to-center distance $dist$ w.r.t.\ the coordinate system depicted in \Fig{Fig:Capsule}.
The dimensions of the box are $576  \, \textup{dx}$ or $768  \, \textup{dx}$ in all three directions, \ie{} $L_x=L_y=L_z$.
For the smaller domain, \num{600000} time steps are performed and for the larger domain \num{605000} time steps.
To keep the system from accelerating infinitely due to the constantly applied force,
the momentum stabilization technique described in \Sect{sec:LBM} is applied.

As fluid, we model water at room temperature, with the kinematic viscosity $\nu_{f,H_2O}$ and the density $\rho_{f,H_2O}$ from \Sect{SubSec:SedimFiberLBMSetup}.
The density of the particles is set to $\rho_p = \SI{1492}{\kilo\gram\per\cubic\meter}$.
The force acting on the particles in z-direction caused by gravity results in the values presented in~\Tab{tab:Forces_tumbling_fluidMatWater}.
\begin{table}[h!]
   \caption{ Gravitational forces applied to the particless of different aspect ratios $1/\eeps$ in z-direction for the tumbling simulations.
            \label{tab:Forces_tumbling_fluidMatWater} }
   \centering
   \begin{tabular}{l lll}
   \toprule
$1/\eeps$              &   10      &   12      &   14     \\
   \midrule
${ F_z} \; [\SI{e-9}{\kilo\gram\meter\per \square\second}]$   &   1.119   &   1.358   &   1.598  \\
   \bottomrule
   \end{tabular}
\end{table}

The resulting particle Reynolds number $\textup{Re}_{p,d}$ for the particles based on the mean sedimentation velocity $U^*$ for the LBM
lies in the range $0.052$ to $0.063$ for  $1/ \eeps = 10$ and $1/ \eeps = 14$, respectively (see~\Tab{tab:SeveralTumblingCapsules_oppVelPeriodLBM}),
so that we are essentially in the Stokes regime.
For the LBM simulations, we apply the relaxation time $\tau=6$ that results in the time increment $\textup{dt} = \SI{4.55e-5}{\second}$ (see~\Eqn{Eq:kinVisc}).\\
In the SBF we use the following parameters $N=5$, $48$ quadrature points along the fiber (see \Eqn{eqn:force_exp}), and $dt \approx 0.003 \, \si{\second}$.

An SBF simulation with $8300$ time steps (TS) was perfomed within $\SI{66.40}{\second}$  on a single core of an Intel Core i7 4770 processor running at $3.4$ GHz,
\ie{}, with \SI{125}{TS\per\second}. 
The parallel runtimes for the LBM on $768$ cores given in \Sect{sec:parallel} correspond to \SI{10.4}{TS\per\second} on LiMa and \SI{3.5}{TS\per\second} on SuperMUC.

\subsection{Flow field around tumbling spherocylinders}
\label{SubSec:TumblingFlowField}
We present the flow field around two tumbling spherocylinders
simulated with the LBM in a periodic domain of size $\left[576 \, \textup{dx} \right]^3$, filled with water.
The setup and parameters described in~\Sect{SubSec:Tumbling_SetupParam} are used
for spherocylinders of aspect ratio $1/\eeps=12$ 
and an initial center-to-center distance of $16 \, \textup{dx}$, with $\textup{dx} = \SI{4.98e-6}{\meter}$.

\begin{figure*}[h!t]
  \centering
  \subfigure[Flow field at beginning of tumbling period \label{subfig:SidewiseMovCaps_tumblingCapsules_Init}]{   
  \includegraphics[bb=0 0 1600 852, trim = 58mm 6mm 217mm 5mm, clip, width=0.41\textwidth]{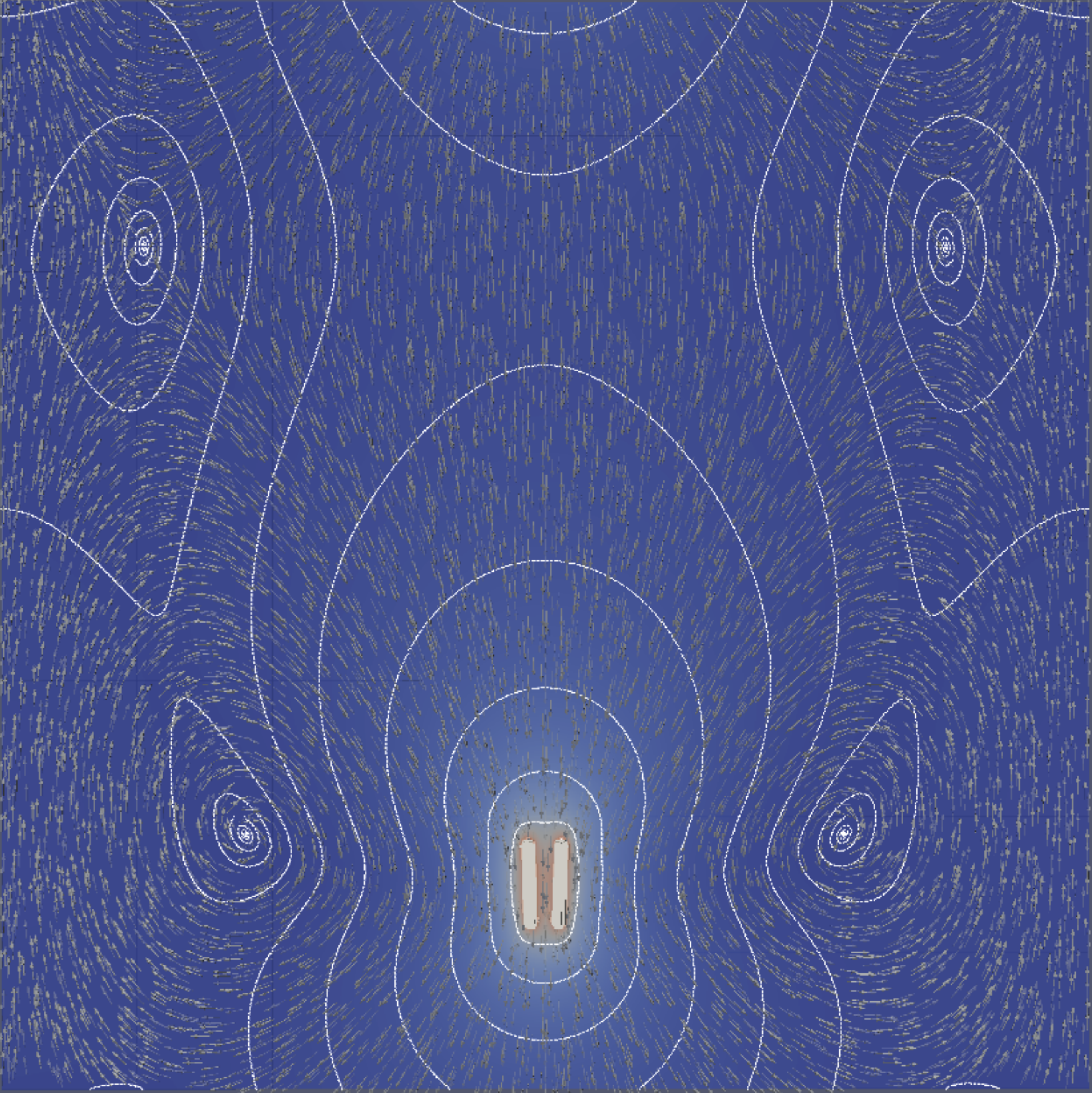}
  }\hspace{5mm}
  \subfigure[Flow field after \num{10000} time steps \label{subfig:SidewiseMovCaps_tumblingCapsules_1000TS}]{
  \includegraphics[bb=0 0 1600 852, trim = 58mm 6mm 217mm 5mm, clip, width=0.41\textwidth]{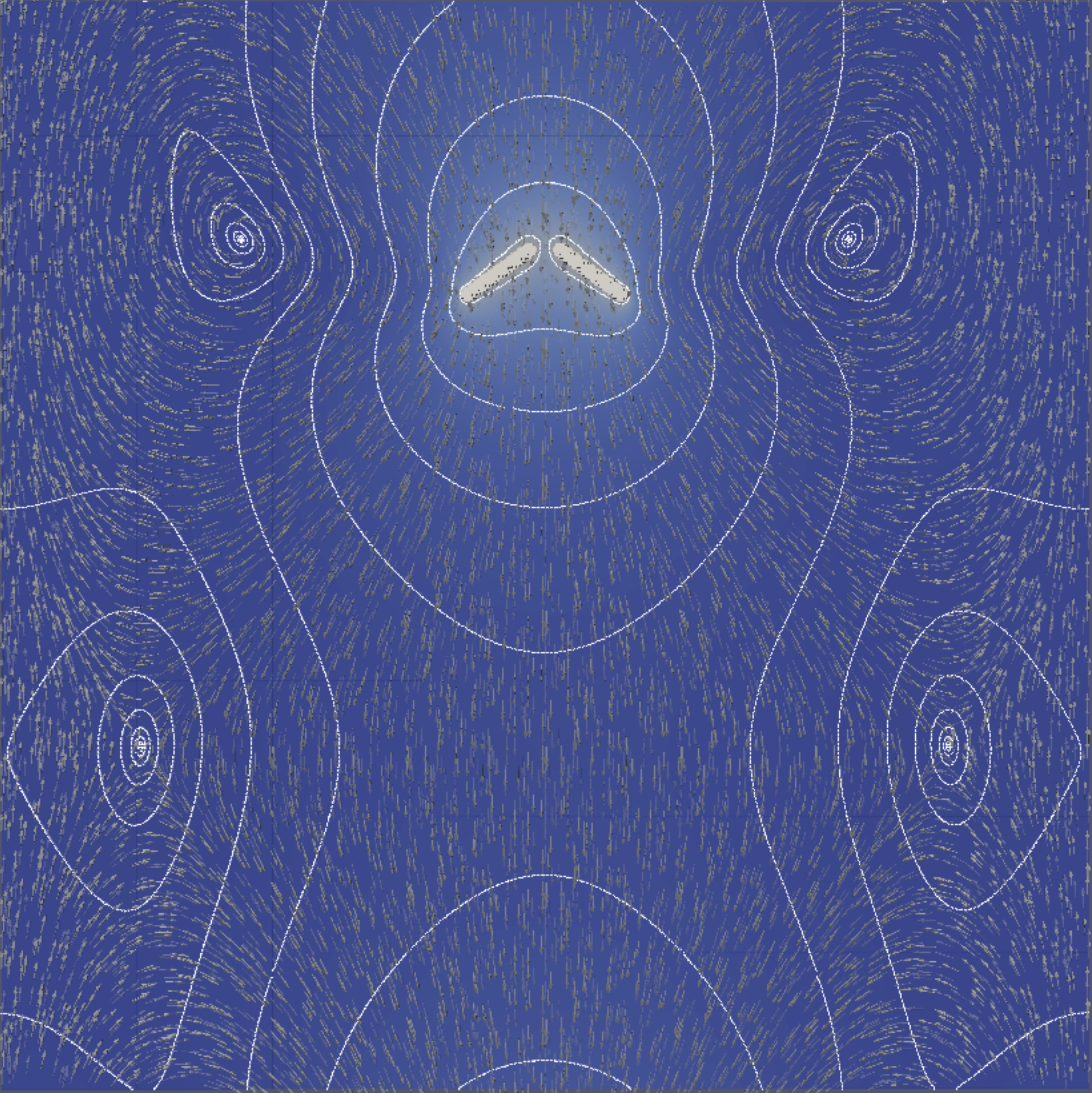}
  }\\
  \subfigure[Flow field after \num{27000} time steps \label{subfig:SidewiseMovCaps_tumblingCapsules_Sidew}]{
  \includegraphics[bb=0 0 1600 852, trim = 58mm 6mm 217mm 5mm, clip, width=0.41\textwidth]{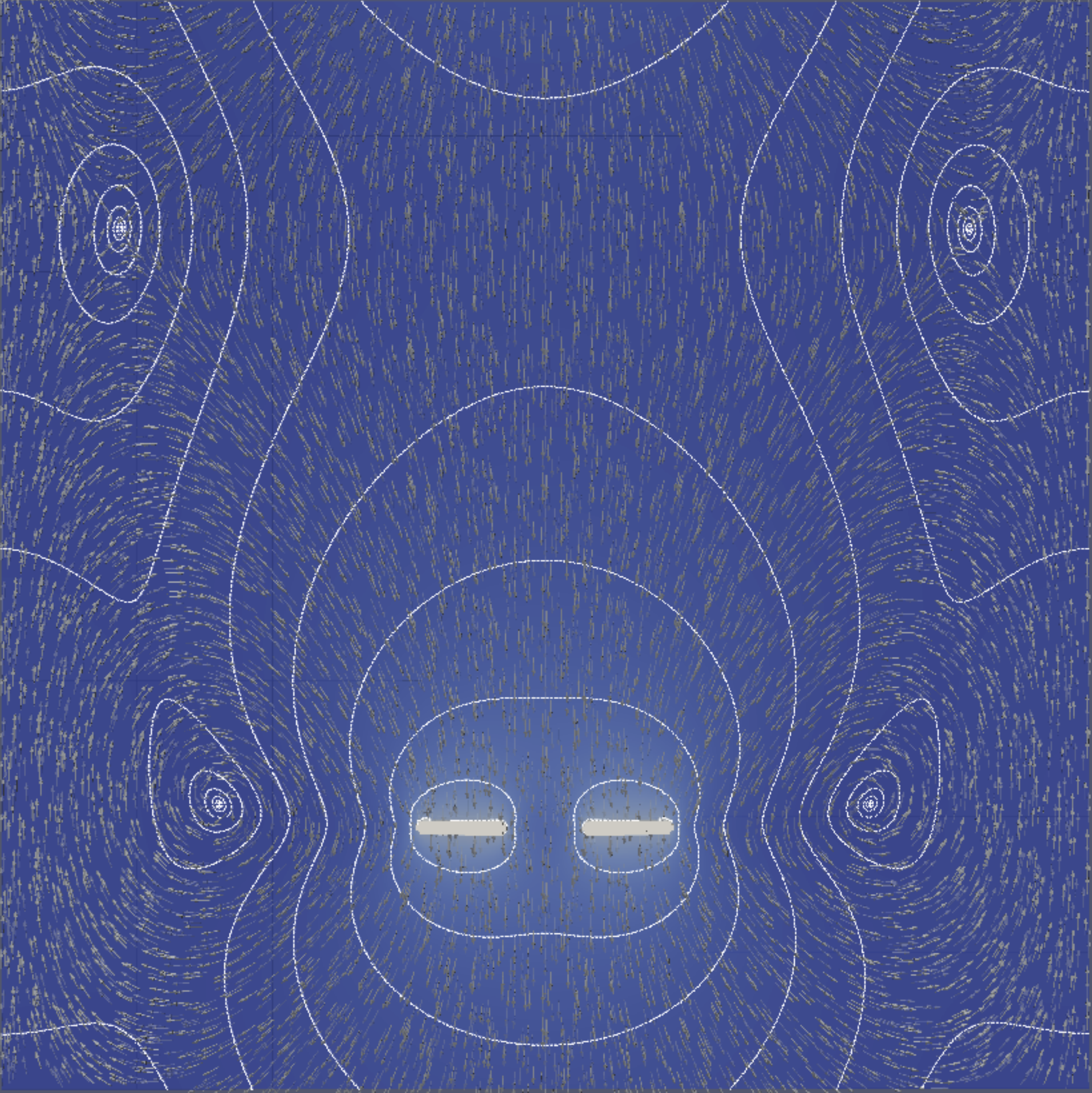}
  }\hspace{5mm}
  \subfigure[Flow field after \num{44000} time steps \label{subfig:SidewiseMovCaps_tumblingCapsules_ScaledGlyphs} ]{
  \includegraphics[bb=0 0 1600 852, trim = 58mm 6mm 217mm 5mm, clip, width=0.41\textwidth]{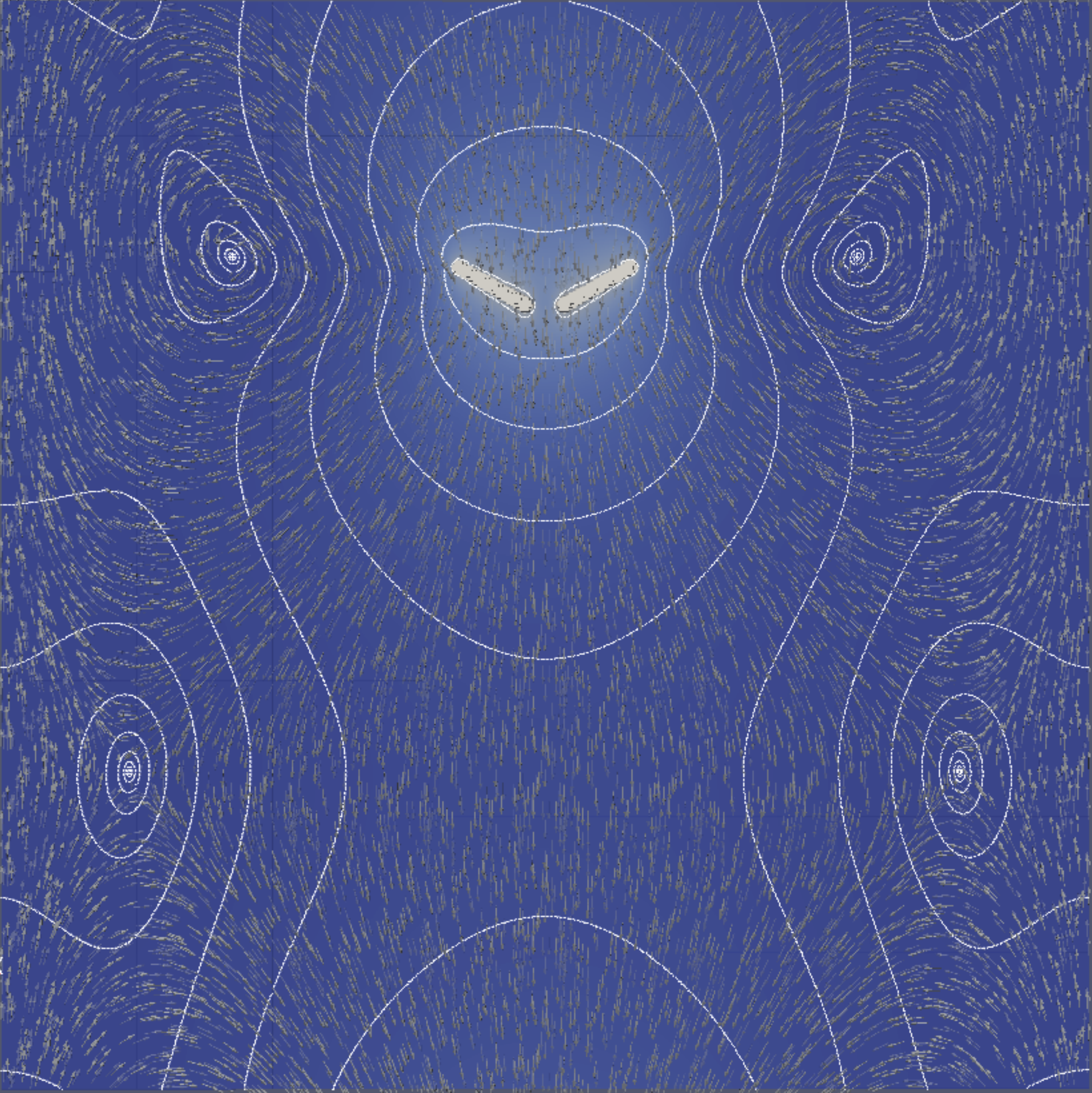}
  }
  \caption{ Flow field around two tumbling spherocylinders with radius $r=4 \textup{dx}$ ($\textup{dx} = \SI{4.98e-6}{\meter}$) 
            and initial distance $16 \, \textup{dx}$ in periodic domain of size $\left[576 \, \textup{dx} \right]^3$.
            Flow direction is indicated by streaks, and velocity magnitude logarithmic contour lines
            and color in range of \SI{3.16e-3}{\meter\per\second} (red) to \SI{20e-9}{\meter\per\second} (blue).
            \label{fig:TumblingCaps_FlowField} }
\end{figure*}
In \Fig{fig:TumblingCaps_FlowField} the flow field is visualized 
along a slice in the x-z plane through the domain center.
The velocity magnitude is represented by a color range from red to blue and by (sixteen) white isosurface contour lines of logarithmic intervals in the range of \SI{2.60e-3}{\meter\per\second} to \SI{96.1e-9}{\meter\per\second}.
The flow direction is represented by streaks of uniform length.
In addition to the image sequence in \Fig{fig:TumblingCaps_FlowField},
an animation is available via a permalink\footnote{\url{https://www10.cs.fau.de/permalink/eethegh4sh}}.\\
At the beginning of the tumbling period, both particles are aligned with the external force direction (see~\Fig{subfig:SidewiseMovCaps_tumblingCapsules_Init}),
and then start rotating (see~\Fig{subfig:SidewiseMovCaps_tumblingCapsules_1000TS}). 
As the particles rotate, they move apart in x-direction until they are oriented perpendicular to the force (see~\Fig{subfig:SidewiseMovCaps_tumblingCapsules_Sidew})
and reach the maximum distance.
The rotation continues, and the particles move further along the force direction (see~\Fig{subfig:SidewiseMovCaps_tumblingCapsules_ScaledGlyphs})
until the next period begins. A more detailed description of the particle positions and velocities is given in \Sect{SubSec:Tumbling_Results}.\\
The fluid is dragged along with the spherocylinders, while the flow velocity quickly decreases with increasing distance from the particles.
As expected, the flow field is at all times symmetric w.r.t.\ the domain center in x- and y-direction and changes in z-direction.
During the tumbling motion, two vortices are forming at the same height (\ie{}, z-coordinate) as the particle center. 
Half-way between periodically following particles in z-direction, a region of zero flow velocity appears at the interface of two vortices
associated with these periodically neighboring particles. 
As the particles change their mutual distance, the vortices and the zero flow velocity regions only slightly change their positions in x-direction.

\subsection{Tumbling results for spherocylinders using LBM}
\label{SubSec:Tumbling_Results}
\begin{figure}[h!]
  \centering
   \includegraphics[bb=0 0 248.47 468.04]{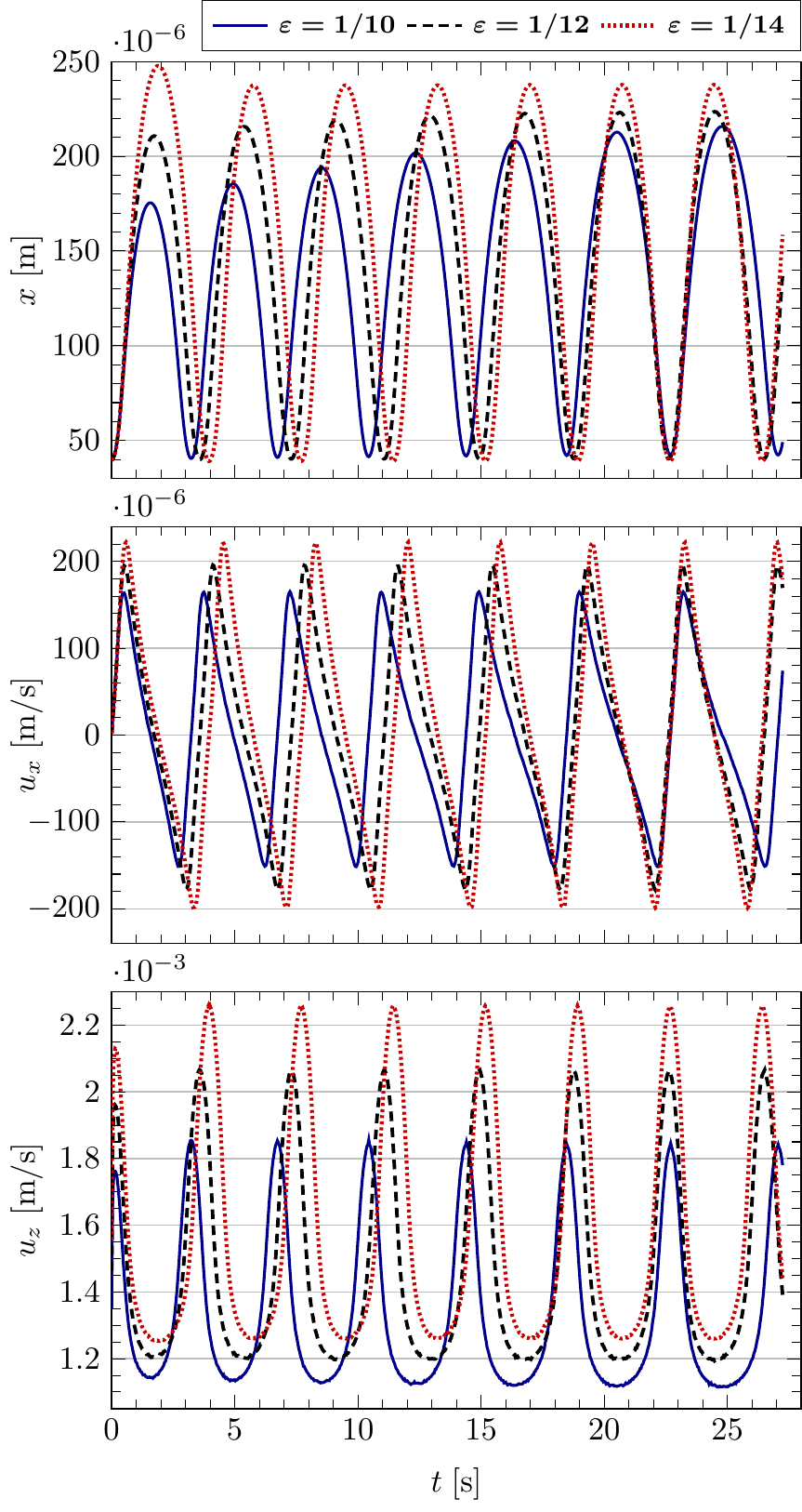} % obtained bounding box required for arxiv after running correctPDFFiles.sh and using values of MediaBox
   \caption{ 
      LBM results for tumbling motion of two spherocylinders with radius $r=4 \, \textup{dx}$ ($\textup{dx} = \SI{4.98e-6}{\meter}$)
      and aspect ratios $1/\eeps=10$, $1/\eeps=12$, and $1/\eeps=14$ 
      in periodic domain of size $[576 \, \textup{dx}]^3$ filled with water.
      Shown are the position of one particle in x-direction and the particle velocities in x- and z-direction over time.
      The spatial discretization is $\textup{dx} = \SI{4.98e-6}{\meter}$, and the initial particle distance is $16 \, \textup{dx}$.
         \label{fig:ResultSeveralTumblingCapsules_DiffAspRatios_Dist8dx_DomSize576} }
\end{figure}

We examine the dependence of the tumbling motion on the spherocylinder aspect ratio, the domain size, and the initial distance between the particles.
The influence of the aspect ratio on the tumbling behavior is analyzed for spherocylinders with $1/\eeps=10$,  $1/\eeps=12$, and $1/\eeps=14$.
For one particle, the position in x-direction and its velocity in x- and z-direction are presented in~\Fig{fig:ResultSeveralTumblingCapsules_DiffAspRatios_Dist8dx_DomSize576} over time.
The simulations are performed for the domain size $[576 \, \textup{dx}]^3$ and the initial center-to-center particle distance of $16 \, \textup{dx}$ in x-direction.

Initially, the particles move away from the domain center and from each other in x-direction.
Once the maximal distance is reached, the particles move back towards each other. This motion is repeated periodically. 
While the particles 
aligned with the direction of gravity move apart, the velocity in x-direction increases, and the particles start rotating.
Before the particles are oriented perpendicular to the gravitational direction, the maximal velocity in x-direction is reached, and the particle velocity in x-direction decreases again.
The velocity in x-direction becomes zero once the particles are oriented perpendicular to gravity.
While the particles continue rotating, the velocity in x-direction becomes negative, \ie{}, the particles approach each other.
The velocity in negative x-direction increases at first and then decreases again as the particles rotate further.
Once the particles are aligned with the direction of gravity, the velocity in x-direction is zero again, and the minimal separation is reached.
In z-direction, the maximal velocity occurs when the particles are oriented with the direction of gravity and the particle distance is minimal.
The minimal velocity in z-direction is reached when the particles are oriented perpendicular to the direction of gravity and the particle distance is maximal.
In both cases, the velocity in x-direction is zero.
The values for the y-direction are not shown since the particles move in the x-z plane, keeping their y-coordinate.

The maximal separation in x-direction, the maximal velocity magnitude in this direction,
and the sedimentation velocity in z-direction all depend on the particle length.
With increasing particle length, the terminal maximal separation distance, the maximal velocity magnitude in x-direction, and the sedimentation velocity $u_z$ increase.
The minimal separation is nearly equal for all particle lengths.
For $1/\eeps=10$, the maximal separation distance per period is small at first and then increases over many periods,
while for $1/\eeps=12$ this distance increases only slightly, and the terminal separation distance is reached after five revolutions.
For $1/\eeps=14$, the separation distance is relatively high in the first period and then decreases until the terminal distance is reached after approximately three revolutions.
Before examining these dependencies further, we analyze how the parameters depend on the initial separation distance between the particles and on the domain size.

\begin{figure}[h!t]
  \centering
   \includegraphics[bb=0 0 248.47 467.56]{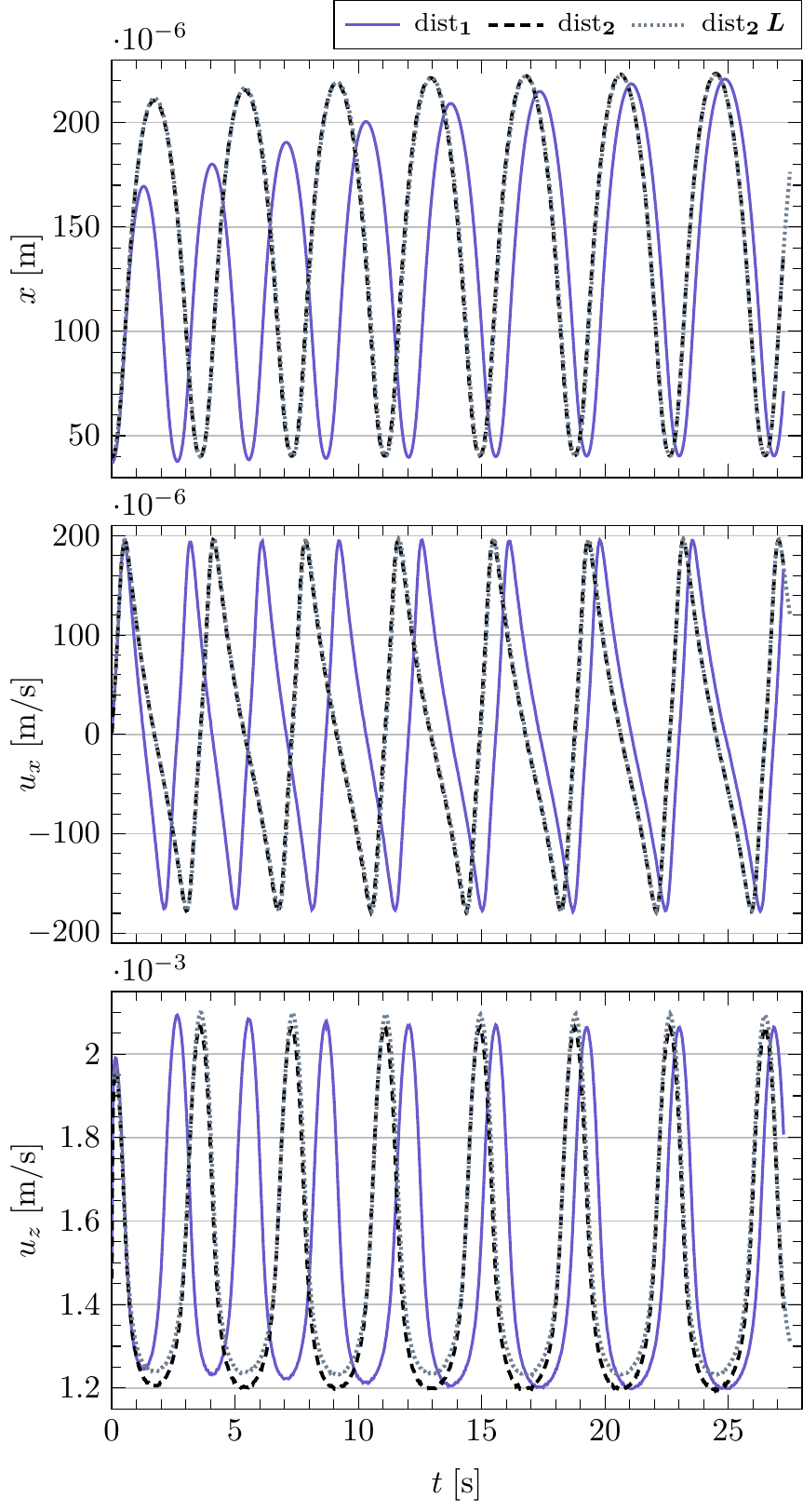}
   \caption{
      LBM results for tumbling motion of two spherocylinders with radius $r=4 \, \textup{dx}$ and aspect ratio $1/\eeps=12$,
      for initial particle distances $\text{dist}_1 = 14.8 \, \textup{dx}$ and $\text{dist}_2 = 16 \, \textup{dx}$
      in water-filled periodic domains of size $\left[576 \, \textup{dx} \right]^3$ and $\left[768 \, \textup{dx} \right]^3$. The larger domain is indicated by `$L$'.}
      \label{fig:ResultSeveralTumblingCapsules_AspRatio12_DiffDistDomSize}
\end{figure}
In \Fig{fig:ResultSeveralTumblingCapsules_AspRatio12_DiffDistDomSize}
we present LBM results for the tumbling motion of particles with aspect ratio $1/\eeps=12$ 
for a smaller initial center-to-center distance of $14.8 \, \textup{dx}$.
Moreover, we compare the particle motion for the previous domain size $\left[576 \, \textup{dx} \right]^3$ and a larger domain of size $\left[768 \, \textup{dx} \right]^3$.
All the other parameters are kept the same as for the previous simulations.
The particle motion conforms to the previously described behavior for the different aspect ratios.\\
The smaller initial distance of the particles leads to a smaller maximal separation at the beginning of the simulation.
Also the tumbling period time is at first smaller than for the higher initial distance.
The velocity in x-direction does not depend on the initial distance, only the sedimentation velocity is a bit higher at first.
After a few periods, however, the tumbling motion for the different initial separations no longer differs.\\
When the domain size is increased from $576 \, \textup{dx}$ to $768 \, \textup{dx}$, the particle motion in x-direction is not affected.
\Ie{}, the size $576 \, \textup{dx}$ suffices for negligible mutual influence of the periodically neighboring particles in x-direction.
Only in z-direction the velocity is smaller for the larger domain, due to the periodic boundaries.

Among the LBM tumbling simulations presented in this section,
the change in separation distance is the highest for $1/\eeps=10$,
and for $1/\eeps=12$ with initial distance $14.8 \, \textup{dx}$.
For these simulations, the measured tumbling period time $T^*$ and period distance $D^*$ in z--direction,
and the resulting mean sedimentation velocity $U^*$ are shown in~\Tab{tab:SeveralTumblingCapsules_CapsDistDepend_TimeEvolution}.
\begin{table}[h!]
   \caption{ Changes in sedimentation distance ($D^*$) per period, period time ($T^*$), and mean sedimentation velocity ($U^*$) for different periods
                 of two tumbling particles with radius $r=4 \, \textup{dx}$ ($\textup{dx} = \SI{4.98e-6}{\meter}$).
                 LBM results for particles with $1/\eeps=10$ and initial distance $16 \, \textup{dx}$, and for $1/\eeps=12$ with initial distance $14.8 \, \textup{dx}$
                 in periodic domain of size $\left[576 \, \textup{dx} \right]^3$.
             \label{tab:SeveralTumblingCapsules_CapsDistDepend_TimeEvolution} }
   \centering
   % \tabsize
   \resizebox{1.04\columnwidth}{!}{
   \begin{tabular}{@{\hspace{0.2ex}}l@{\hspace{0.8ex}} l ccccccc}
   \toprule
     $1/\eeps$                 &  period \#                                         & 1      &  2     &  3     &  4     &  5     &  6     & 7      \\  
   \cmidrule(r){1-1}  \cmidrule(l r){2-9} 
   \multirow{3}{*}{$10$} & $D^* \, [\SI{e-3}{\meter}]$   &  4.38  &  4.78  &  4.97  &  5.25  &  5.41  &  5.54  &  5.72  \\  
                               & $T^* \, [\si{\second}]$                       &  2.71  &  3.52  &  3.69  &  3.94  &  4.09  &  4.23  &  4.35  \\ 
                               & $U^* \, [\SI{e-3}{\meter\per\second}]$ &  1.61  &  1.36  &  1.35  &  1.33  &  1.32  &  1.31  &  1.31  \\  
   \cmidrule(r){1-1}   \cmidrule(r){2-2} \cmidrule(l r){3-9}  
   \multirow{3}{*}{$12$ } & $D^* \, [\SI{e-3}{\meter}]$  &  4.18  &  4.50  &  4.75  &  5.02  &  5.24  &  5.42  &  5.51  \\  
                               & $T^* \, [\si{\second}]$                       &  2.13  &  2.92  &  3.08  &  3.34  &  3.53  &  3.70  &  3.76  \\  
                               & $U^* \, [\SI{e-3}{\meter\per\second}]$ &  1.96  &  1.54  &  1.54  &  1.50  &  1.49  &  1.47  &  1.47  \\  
   \bottomrule
   \end{tabular}
   }
\end{table}
For both aspect ratios, the tumbling period distance increases as the particles move apart in x-direction.
The tumbling period time increases at almost the same rate, leading to an only slightly decreasing mean sedimentation velocity in both cases.
The terminal sedimentation velocity for $1/\eeps=12$ is higher than for $1/\eeps=10$.

The main characteristic parameters for the tumbling motion of two spherocylinders obtained from the LBM simulations are summarized in~\Tab{tab:SeveralTumblingCapsules_oppVelPeriodLBM}.
\begin{table}[h!b]
   \caption{ Parameters obtained from LBM simulations of two tumbling particles with radius $r=4 \textup{dx}$ ($\textup{dx} = \SI{4.98e-6}{\meter}$) in periodic cubic domain filled with water.
                Sedimentation distance ($D^*$),  period time ($T^*$), and mean velocity ($U^*$) in z-direction,
                minimum and maximum velocities in z-direction ($u_\text{z, min}$, $u_\text{z, max}$) and x-direction ($u_\text{x, min}$, $u_\text{x, max}$)  per revolution,
                are shown for different aspect ratios ($1/\eeps$), domain sizes ($L_{x,y,z}$), and initial particle distances ($dist$).
                 \label{tab:SeveralTumblingCapsules_oppVelPeriodLBM} }
   \centering
   % \tabsize
   \begin{tabular}{l@{\hspace{0.7ex}} l ccccc}
   \toprule
      $1/\eeps$         &                                   & 10     &  12    &  12    &  14    &  12    \\
      \cmidrule(r){1-2}                                    \cmidrule(lr){3-6}                  \cmidrule(lr){7-7}
      $L_{x,y,z}$       &  {[}\textup{dx}{]}                &  576   &  576   &  576   &  576   &  768   \\
      $dist$            &  {[}\textup{dx}{]}                &  16.0  &  14.8  &  16.0  &  16.0  &  16.0  \\
      \cmidrule(r){1-2}                                    \cmidrule(lr){3-6}                  \cmidrule(lr){7-7}
      $D^*$             &  $[\SI{e-3}{\meter}]$             &  5.72  &  5.63  &  5.63  &  5.92  &  5.71  \\
      $T^*$             &  $[\si{\second}]$                 &  4.35  &  3.82  &  3.86  &  3.75  &  3.85  \\
      $U^*$             &  $[\SI{e-3}{\meter\per\second}]$  &  1.31  &  1.47  &  1.46  &  1.58  &  1.48  \\
      \cmidrule(r){1-2}                                    \cmidrule(lr){3-6}                  \cmidrule(lr){7-7}
      $u_\text{z, min}$ &  $[\SI{e-3}{\meter\per\second}]$  &  1.11  &  1.20  &  1.19  &  1.25  &  1.23  \\
      $u_\text{z, max}$ &  $[\SI{e-3}{\meter\per\second}]$  &  1.86  &  2.09  &  2.07  &  2.26  &  2.10  \\
      \cmidrule(r){1-2}                                    \cmidrule(lr){3-6}                  \cmidrule(lr){7-7}
      $u_\text{x, min}$ &  $[\SI{e-6}{\meter\per\second}]$  &  -152  &  -179  &  -178  &  -199  &  -177  \\
      $u_\text{x, max}$ &  $[\SI{e-6}{\meter\per\second}]$  &  164   &  196   &  196   &  222   &  196   \\
   \bottomrule
   \end{tabular}
\end{table}
The terminal mean sedimentation velocity $U^*$ per period rises with increasing aspect ratio or particle length,
which agrees with the experimental results in~\cite{JungSedim2006PhysRevE.74.035302}.
This behavior is contrary to the sedimentation results in \Sect{sec:comparing}, where the same force is applied to all particles
and can thus be attributed to the increasing gravitational force acting on the particles.
For the larger domain, $U^*$ increases slightly, as observed previously for $u_z$.
While the terminal mean sedimentation velocity is reached for all simulations,
the tumbling period distance $D^*$ and time $T^*$ still change for $1/\eeps=10$ and for $1/\eeps=12$ with $dist=14.8 \, \textup{dx}$
(see also \Tab{tab:SeveralTumblingCapsules_CapsDistDepend_TimeEvolution}).
For the latter, the values converge towards the results for $dist=16 \, \textup{dx}$.
The period time decreases with increasing aspect ratio, whereas no clear trend is recognizable for $D^*$.
For the larger domain and $1/\eeps=12$, the smaller sedimentation velocity results mainly from an increased distance $D^*$.\\
The velocities in x-direction are significantly smaller than the sedimentation velocities.
As reported above, the maximal velocity magnitudes in x-direction increase with particle length.
However, these velocities do not depend on the domain size and the initial separation distance.
The maximal magnitude $u_\text{x, max}$ when the particles separate is higher than the magnitude $u_\text{x, min}$ for the approaching particles.
Moreover, the maximal distance between the particles in x-direction converges to a value that depends on the aspect ratio, 
but not on the initial separation (see \Fig{fig:ResultSeveralTumblingCapsules_DiffAspRatios_Dist8dx_DomSize576} and \Fig{fig:ResultSeveralTumblingCapsules_AspRatio12_DiffDistDomSize}). 
The minimal distance is hardly affected by these parameters.

\subsection{Tumbling results for ellipsoids using SBF and comparison to LBM results}
The slender body formulation results for two tumbling particles in a periodic domain are presented in \Tab{tab:SeveralTumblingCapsules_oppVelPeriodSBF}
for different domain sizes and particle distances equivalent to the LBM simulations.
\begin{table}[h!]
   \caption{   Simulation results using the slender body formulation
               for two tumbling fibers  in a periodic domain filled with water.
               The spherocylinder radius is $r=4 \textup{dx}$. \label{tab:SeveralTumblingCapsules_oppVelPeriodSBF} }
   \centering
   \begin{tabular}{l@{\hspace{0.7ex}} l ccccc}
   \toprule
      $1/\eeps$         &                                   &  10      &   12      &   12      &   14      &   12      \\
   \cmidrule(r){1-2}                                    \cmidrule(lr){3-6}                  \cmidrule(lr){7-7}
      $L_{x,y,z}$       &   {[}\textup{dx}{]}               &  576     &   576     &   576     &   576     &   768     \\
      $dist$            &   {[}\textup{dx}{]}             &  15.0   &   14.9   &   15.1   &   15.1   &   15.1   \\
   \cmidrule(r){1-2}                                    \cmidrule(lr){3-6}                  \cmidrule(lr){7-7}
      $D^*$             &   $[\SI{e-3}{\meter}]$            &  6.40    &   6.17    &   7.20    &   9.01    &   5.31    \\
      $T^*$             &   $[\si{\second}]$                &  4.50    &   3.91    &   4.53    &   5.73    &   3.84    \\
      $U^*$             &   $[\SI{e-3}{\meter\per\second}]$ &  1.42    &   1.57    &   1.53    &   1.58    &   1.61    \\
   \cmidrule(r){1-2}                                    \cmidrule(lr){3-6}                  \cmidrule(lr){7-7}
      $u_\text{z, min}$ &   $[\SI{e-3}{\meter\per\second}]$ &  1.23    &   1.32    &   1.31    &   1.35    &   1.36    \\
      $u_\text{z, max}$ &   $[\SI{e-3}{\meter\per\second}]$ &  2.00    &   2.36    &   2.35    &   2.58    &   2.40    \\
   \cmidrule(r){1-2}                                    \cmidrule(lr){3-6}                  \cmidrule(lr){7-7}
      $u_\text{x, min}$ &   $[\SI{e-6}{\meter\per\second}]$ &  -180   &   -224   &   -224   &   -260   &   -224   \\
      $u_\text{x, max}$ &   $[\SI{e-6}{\meter\per\second}]$ &  180    &   224    &   224  &   260    &   224    \\
   \bottomrule
   \end{tabular}
\end{table}
Similar to the LBM results, both $u_z$ and $U^*$ increase with the aspect ratio and with the domain size for constant aspect ratio.
However, these velocities are more sensitive to the domain size than for the LBM.
Moreover, the initial distance influences $U^*$ stronger than for the LBM, whereas $u_z$ is hardly affected.
In contrast to the LBM results, the SBF period time $T^*$ does not systematically depend on the aspect ratio
whereas the period distance $D^*$ tends to increase with $1/\eeps$. However, $D^*$ depends strongly on both domain size and initial separation distance.
With increasing aspect ratio, the velocities $u_x$ increase but are not sensitive to initial distances and domain sizes, like for the LBM.

A comparison of the values presented in \Tab{tab:SeveralTumblingCapsules_oppVelPeriodSBF} except for $T^*$ and $U^*$
to the corresponding values in \Tab{tab:SeveralTumblingCapsules_oppVelPeriodLBM}
reveals a relative difference of approximately $10-20 \%$ for
$1/\eeps=10, 12$. This difference can be attributed to the
different shapes of the particles in the LBM and the SBF
respectively, see the discussion in \Sect{SubSec:SedimFiberResults}.
However, for $1/\eeps=14$ the difference is larger, especially for
quantities related to the periodicity of the tumbling motion, $T^*$ and 
$D^*$. For these quantities the relative difference is now in the order of $35 \%$.
In addition to the effect of different particle shape,
the combined effect of particle and fluid inertia
could play a role here that are both only present in the LBM.
Since the velocity of the tumbling
particles increases with growing particle lengths, so does the
Reynolds number as well as the importance of fluid inertia.\\
Furthermore, all values obtained with the LBM simulations are
consistently lower than the values obtained with the SBF. All of the
above is in line with the findings for one sedimenting particle
presented in \Sect{SubSec:SedimFiberResults}.

\begin{figure}[h!]
  \centering
   \includegraphics[bb=0 0 248.47 467.55]{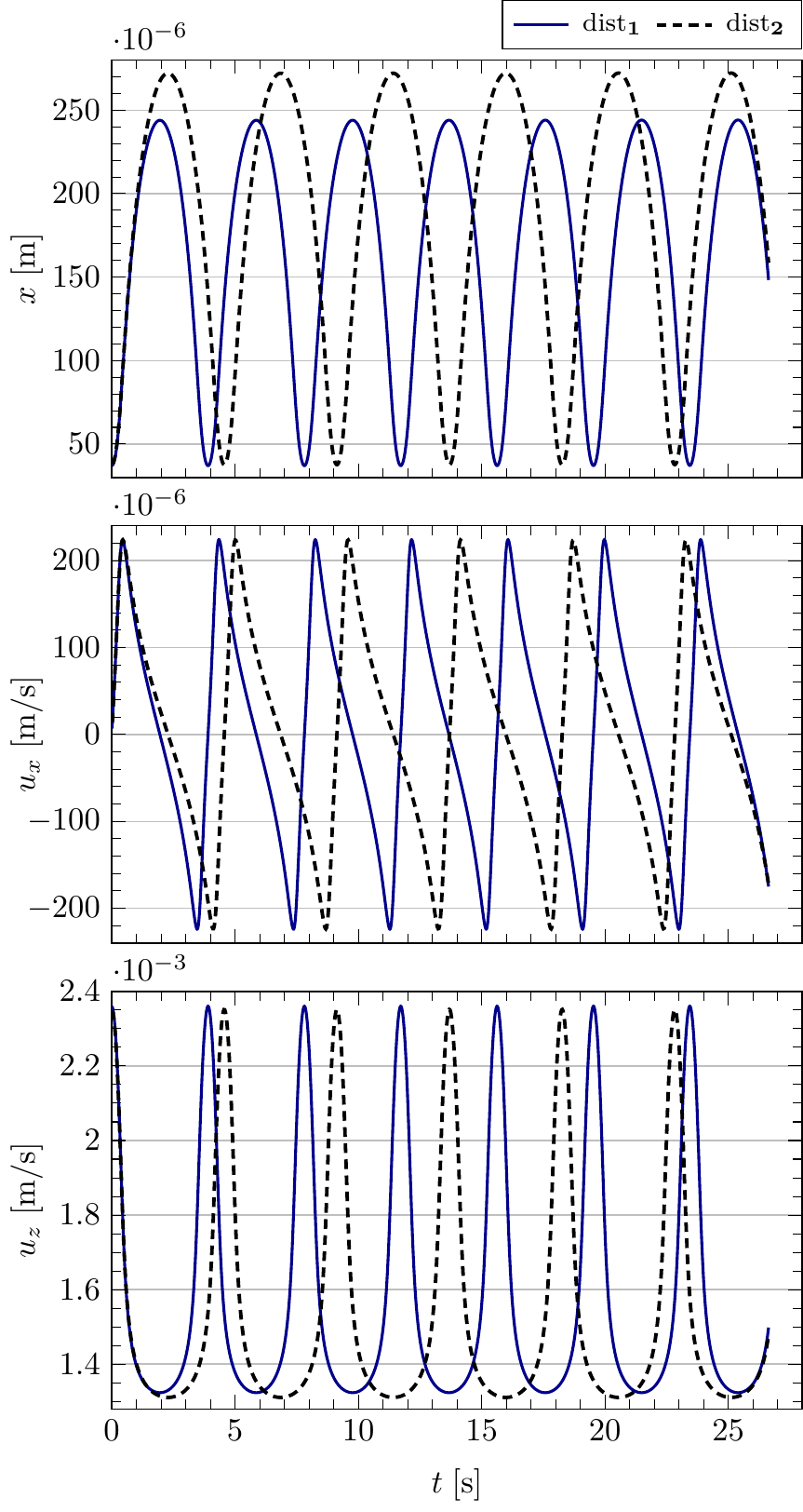}
   \caption{ 
      SBF results for tumbling motion of two fibers with radius $r=4 \, \textup{dx}$ 
      and aspect ratios $1/\eeps=12$ for initial fiber distances of $\text{dist}_1 = 14.9 \, \textup{dx}$ and $\text{dist}_2 = 15.1 \, \textup{dx}$. 
      The size of the  periodic domain is  $[576 \, \textup{dx}]^3$.
      \label{fig:SBF2_Tumbl} }
\end{figure}

Another difference between the two methods that also
can be attributed to
inertial effects only being present in the LBM simulations, is that
the SBF results depend strongly on the initial
center-to-center distance between the particles, see \Fig{fig:SBF2_Tumbl}.
For the LBM the situation is different.
In \Fig{fig:ResultSeveralTumblingCapsules_AspRatio12_DiffDistDomSize}
we see that the difference caused by starting with two particles at
two different center-to-center distances decreases with time, and the
curves will eventually coincide. In the SBF, the difference in the
resulting curves stay the same throughout the whole simulation.

In the upper diagram of \Fig{fig:LBMSBFTumblComparison} we show a
comparison of the sedimentation velocity for a particle with $1/\eeps=12$
obtained with the LBM and the SBF for initial distances $16\,\textup{dx}$ and $14.9\,\textup{dx}$, respectively. 
The results agree well on a qualitative level.  
However, for the SBF the sedimentation velocity is larger and
the period time is slightly longer than for the LBM. This is
consistent with the results for single particles presented in \Sect{SubSec:SedimFiberResults},
where the particle velocities of the SBF for ellipsoids are higher than those of the LBM for spherocylinders,
due to the different particle shapes.

\subsection{Comparison of tumbling orbits from LBM and SBF}
The tumbling motion of elongated particles can also be characterized by orbits
in phase space, \ie{}, the space of particle positions and velocities.
\begin{figure}[h!b]
  \centering
\begin{tabular}{l}
   \includegraphics[bb=0 0 248.47 179.44]{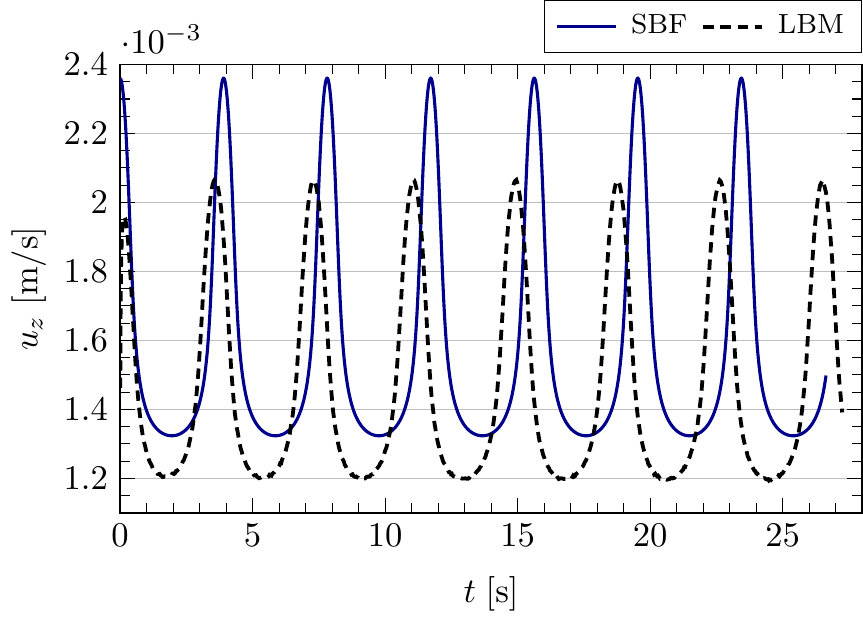}\\  
   \includegraphics[bb=0 0 252.59 336.93]{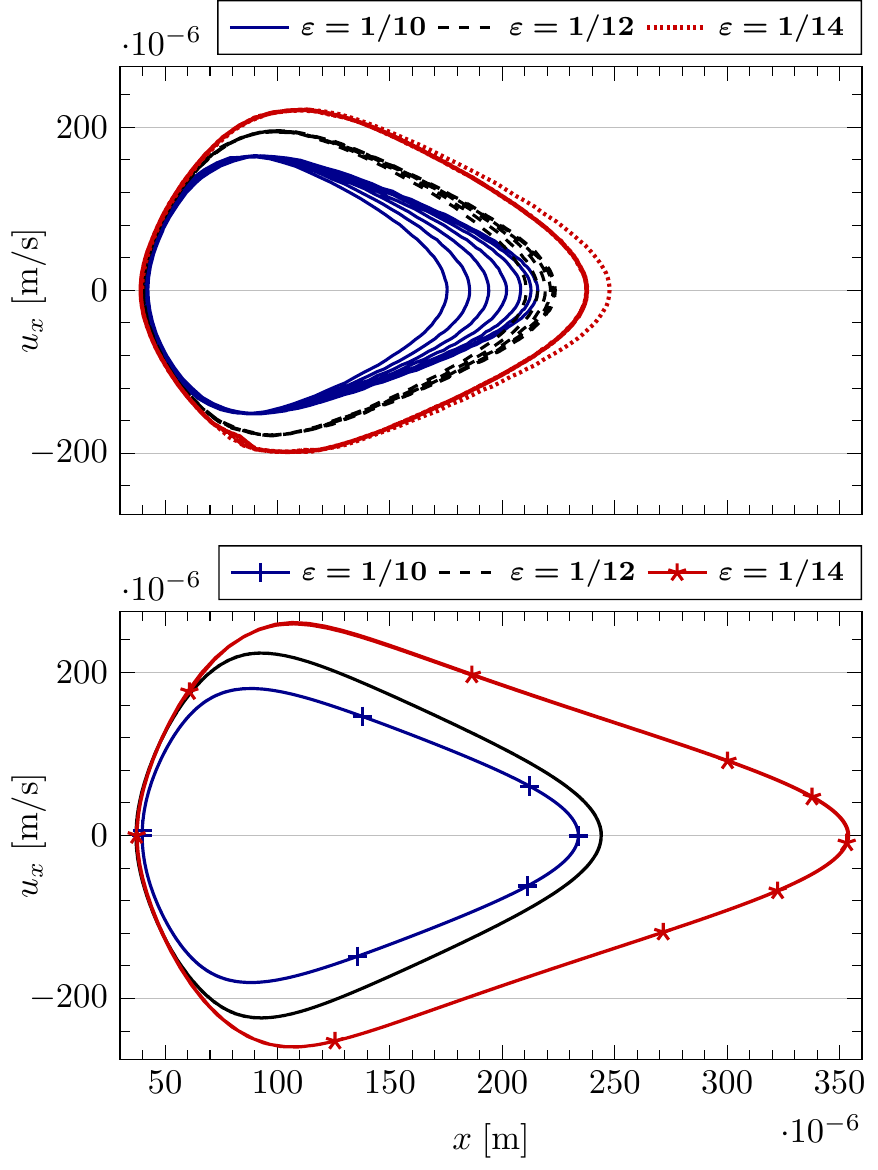}
\end{tabular}
  \caption{ Comparison of the results from LBM and SFB for tumbling motion of two particles with radius $r=4 \, \textup{dx}$
            and different aspect ratios $1/\eeps$ and initial distance $16 \, \textup{dx}$ in a periodic domain of size $[576 \, \textup{dx}]^3$. % filled with water.
            The upper diagram shows the particle velocity in z-direction over time for $1/\eeps=12$.
            The middle and lower diagrams show the tumbling orbits in phase-space (x-position vs. x-velocity) for LBM and SBF, respectively.
            \label{fig:LBMSBFTumblComparison}
          }
\end{figure}
Since the particle motion is periodic in the direction perpendicular
to the sedimentation direction, the orbits in phase space are
represented by closed curves. We present the periodic orbits for LBM
and SBF simulations.

The tumbling orbits obtained from the LBM simulations are shown in the
middle diagram of \Fig{fig:LBMSBFTumblComparison} for spherocylinders 
in a periodic, cubic domain. As shown in
\Fig{fig:ResultSeveralTumblingCapsules_DiffAspRatios_Dist8dx_DomSize576},
the LBM simulations converge towards periodic motion in x-direction
after a few revolutions, \ie{}, the orbits are asymptotically periodic.
This effect can be attributed to the inertia considered by the LBM.
For the aspect ratio $1/\eeps = 10$, the convergence takes
the longest, with increasingly larger cycles in x-direction.
The reason for this convergence behavior is that the initial separation distance for $1/\eeps = 10$ 
differs the most from the preferred distance (see \Fig{fig:ResultSeveralTumblingCapsules_DiffAspRatios_Dist8dx_DomSize576}).
The next longer particles with $1/\eeps = 12$ converge faster, and the maximal
separation in x-direction gets larger at first, too. For $1/\eeps = 14$, 
the maximal particle separation at the
first revolution is the highest but converges quickly. 
Both, terminal maximal separation and velocity magnitude, increase
with particle length. However, minimal positions and maximal velocity magnitudes 
in x-direction do not change with the revolutions. 
Moreover, the orbits are not symmetric w.r.t.\ the
ordinate because the maximal separation velocity is higher than the
maximal approach velocity.

The periodic orbits obtained from the SBF simulations are shown in the
lower diagram in \Fig{fig:LBMSBFTumblComparison}. These orbits exhibit
in principle the same behavior as the LBM results. The shape of the
curves is similar, and the maximal separation distance increases with
the particle length. However, the initial minimum separation distance
is retained and thus differs for the aspect ratio $1/\eeps=10$ from
the longer particles. Moreover, the terminal periodic orbit is already
obtained after the first revolution as there is no inertia, and the maximal separation and
approach velocities have the same magnitude for a given particle length.%
\section{Conclusions	}
\label{sec:conclusions}
In this article, we have validated and compared different methods for elongated particles in creeping flows.
The SBF models hydrodynamic interactions of elongated ellipsoidal particles in flow based on an asymptotic formulation in the slenderness parameter.
Hence the model accuracy increases with aspect ratio.
The LBM simulates hydrodynamic interactions by an explicit computation of the flow field and models the momentum exchange between fluid and particles.
Here, the flow is represented on a lattice, and the particles are modeled as spherocylinders.

% \paragraph{Single particle motion}~\\
To validate the LBM simulations, LBM and SBF results for the translational and rotational motion of single elongated particles
were compared to analytical solutions for cylinders according to the theories of Cox~\cite{Cox70} and Tirado et~al.~\cite{tirado1984comparison}.
While the different theories for cylinders and the results for spherocylinders converge quickly with increasing aspect ratio $1/\eeps$,
the SBF solutions for ellipsoids approach these results only for extremely high aspect ratios.
The particle shape is identified as the dominant factor for the different velocities of (sphero-)cylinders and ellipsoids.
This shape effect explains the consistently lower velocities for the LBM than for the SBF that deviate by about $10 \%$ to $16\%$ for translation and by about $12 \%$ to $13 \%$ for rotation at $1/\eeps \geq 8$.
Overall, the terminal velocities of the different theories converge faster for sidewise orientation w.r.t.\ the applied force than for lengthwise orientation.
The translational velocities for spherocylinders agree very well with the theory of Tirado~et~al.\ for cylinders of full spherocylinder lengths,
apart from retarding wall effects in the LBM simulations at high aspect ratios.
For rotational motion, the angular velocities of spherocylinders are slightly higher than predicted by Tirado~et~al.\ for cylinders of full spherocylinder length.
We find that in the LBM simulations of single particles, inertial effects play only a minor role compared to shape and wall effects.
Finally, the flow field around sidewise and lengthwise moving spherocylinders obtained from LBM simulations is compared.
The comparison shows that for sidewise motion the fluid is affected at higher distances, leading to lower sedimentation velocities.
Moreover, the counterflux of fluid in the closed domain becomes visible that results in the retarding wall effect.

% \paragraph{Tumbling motion}~\\
After the single particle motion validation, the tumbling motion of two elongated particles was examined in a periodic domain.
For the LBM simulations, the flow field around the tumbling particles was visualized.
The particle trajectories obtained from both, LBM and SBF simulations, were compared in diagrams, 
including a representation of the tumbling orbits in phase-space.
The qualitative agreement of the LBM and SBF simulation results is good, whereas quantitative differences become apparent.
Consistent with the single particle motion, the characteristic tumbling parameters differ by about $10-20 \%$ for aspect ratios of $1/\eeps=10, 12$. 
For an aspect ratio of 14 the differences increase, which might partially be attributed to higher differences in the velocities of single lengthwise sedimenting particles. 
~\\
In contrast to the SBF, a preferred maximal particle separation distance exists for the LBM in the direction perpendicular to gravity,
which depends on the aspect ratio and can be attributed to inertia.
This effect is similar to the drift to stable orbits due to inertia examined in Mao \& Alexeev~\cite{mao2014motion}
for the motion of spheroidal particles in shear flow.
Moreover, Jung et~al.~\cite{JungSedim2006PhysRevE.74.035302} report that the experimental tumbling results for rods are insensitve to the initial separation and converge to the preferred distance.
Also the increasing mean sedimentation velocities with particle length observed in the LBM and SBF simulations agree with the experimental investigations in~\cite{JungSedim2006PhysRevE.74.035302}. 

% \paragraph{Stengths/Weaknesses of LBM and SBF}~\\
The strengths of the SBF compared to the LBM lie in 
its significantly lower computational effort as well as in its ability to simulate zero Reynolds number flows.
However, the SBF relies on high aspect ratios to accurately model elongated particles,
and other boundary conditions than periodic or free space are not as straight forward as in the LBM.
The LBM can in contrast 
simulate complex geometries and nearly arbitrary particle shapes, only restricted
by the \pe{} in this respect.
Moreover, the LBM is able to consider inertia, which allows physically more realistic simulations and a better comparison to experiments.
The high parallel efficiency of the LBM allows its execution on massively parallel clusters and compensates for its higher computational effort.

% \paragraph{Model extensions}~\\
A possible extension of the SBF is the implementation of wall boundary conditions.
Wall treatments in a boundary integral setting have been found to be
challenging, and the literature is sparse. The most straightforward way
to include these outer boundaries is to treat the wall in the same way as the immersed objects. 
A boundary integral over the wall is incorporated in the formulation and 
is discretized using special quadrature.
This approach yields extra
unknowns that must be computed
and thus increases computational cost.
The number of additional unknowns will depend on the size of the domain as well
as on the resolution.
A different way of including wall boundary conditions in the SBF 
is to use the method of images\cite{Blake71}. 
However, this method is only feasible
for one or two parallel plane boundaries.

% \paragraph{Future work}~\\
In the future we may compare the two examined methods for elongated particles in flows subject to wall boundary conditions
and simulate the interactions of many sedimenting particles, including LBM simulations at higher Reynolds number.
The LBM may also be applied to simulate collisions of elongated particles with walls and their motion in complex geometries.
Further work could include multiphysics simulations of
charged elongated particles in fluids subject to electric fields
and their deposition on a charged surface,
similar to the simulations with spheres in~\cite{Bartuschat:2014:CP}.%

\begin{appendix}
  \section{Detailed LBM results for single particle motion\label{sec:AppendixSPM}}
  Additional details on LBM results from \Sect{sec:comparing}.
\begin{table}[h!t]
  \caption{ LBM sedimentation velocities $U^{*}_\text{LBM}$ of spherocylinders with different aspect ratios $1/\eeps$ % (or lengths $L$)
                 and radius $r=4 \, \textup{dx}$ ($\textup{dx} = \SI{e-5}{\meter}$),
                 in $[2560 \, \textup{dx}]^2 \times 2688 \, \textup{dx}$  domain with free-slip boundaries,
                 and angular velocities $\omega^{*}_\text{LBM}$ in $[816 \, \textup{dx}]^3$ domain with no-slip boundaries.
                 Fluctuations due to obstacle mappings $\delta_\text{U} = ({U^*_\text{max}-U^*_\text{min}})/{U^*_\text{LBM}}$ and $\delta_\omega$ (defined analogeously) are shown, 
                 together with Reynolds numbers based on $U^{*}_\text{LBM}$ and diameter ($\textup{Re}_{p,d}$) or length ($\textup{Re}_{p,L}$), 
                 or tip velocity ($\textup{Re}_{\text{tip},d}$, $\textup{Re}_{\text{tip},L}$).
                 LBM velocities for lengthwise/sidewise translational-- and rotational motion are compared to
                 analytical solutions by Tirado et~al.~\cite{tirado1984comparison} for cylinders with lengths including  (indicated by `Tir,wC')  
                 and excluding (`Tir,nC') spherocylinder end-caps.
                 Differences between LBM and Tirado solutions are shown as relative deviations $\Delta_\text{r}U$ and $\Delta_\text{r}\omega$ w.r.t.\ the Tirado solutions.
                 \label{tab:TerminalSedimVel_Deviation_aspRatio_2560Domain_LBM_fluidMatWater} }
      \centering
\resizebox{1.05\linewidth}{!}{%
   \begin{tabular}{ |c|l@{\hspace{0.2ex}}l@{\hspace{0.5ex}}|llllll|}
\hline
               &   $1/\eeps$      &                                  &   4          &   6          &   8          &   10         &   12         &   14         \\
               &   $L$& $[\SI{e-3}{\meter}]$             &   0.16       &   0.24       &   0.32       &   0.40       &   0.48       &   0.56       \\
\hline
\hline
\multirow{8}{*}{\parbox{0.8mm}{l\\e\\n\\g\\t\\h\\w}} &
                   $U^{*||}_\text{LBM}$&$[\SI{e-6}{\meter\per\second}]$&   503        &   419        &   363        &   324        &   294        &   269        \\
               &   $\delta_\text{U}$   &\,[\%]                       &   0.80       &   0.47       &   0.15       &   0.40       &   0.35       &   0.30       \\
               &   $\textup{Re}_{p,d}$ &                             &   0.040      &   0.034      &   0.029      &   0.026      &   0.023      &   0.022      \\
               &   $\textup{Re}_{p,L}$ &                 &   0.080      &   0.10       &   0.12       &   0.13       &   0.14       &   0.15       \\
\cline{2-9}
               &   $U_\text{Tir,nC}$&$[\SI{e-6}{\meter\per\second}]$ &   653        &   481        &   409        &   361        &   325        &   297        \\
               &   $\Delta_\text{r}U$&\,[\%]                          &   -22.9      &   -12.9      &   -11.2      &   -10.2      &   -9.7       &   -9.3       \\
\cline{2-9}
               &   $U_\text{Tir,wC}$&$[\SI{e-6}{\meter\per\second}]$ &   481        &   409        &   361        &   325        &   297        &   273        \\
               &   $\Delta_\text{r} U$&[\%]                         &   4.59       &   2.38       &   0.65       &   -0.28      &   -1.05      &   -1.58       \\
\hline
\hline
\multirow{8}{*}{\parbox{0.8mm}{s\\i\\d\\e\\w\\i\\s\\e}} &
                   $U^{*\bot}_\text{LBM}$&$[\SI{e-6}{\meter\per\second}]$              &   447        &   349        &   291        &   252        &   223        &   201        \\
               &   $\delta_\text{U}$&[\%]                           &   0.78       &   0.80       &   0.79       &   0.87       &   0.83        &   0.81        \\
               &   $\textup{Re}_{p,d}$&                              &   0.036      &   0.028      &   0.023      &   0.020      &   0.018      &   0.016      \\
               &   $\textup{Re}_{p,L}$&                  &   0.07       &   0.08       &   0.09       &   0.10       &   0.11       &   0.11       \\
\cline{2-9}
               &   $U_\text{Tir,nC}$&$[\SI{e-6}{\meter\per\second}]$ &   641        &   429        &   344        &   292        &   255        &   227        \\
               &   $\Delta_\text{r}U$&[\%]                          &   -30.3      &   -18.6      &   -15.4      &   -13.6      &   -12.5      &   -11.6      \\
\cline{2-9}
               &   $U_\text{Tir,wC}$&$[\SI{e-6}{\meter\per\second}]$ &   429        &   344        &   292        &   255        &   227        &   205        \\
               &   $\Delta_\text{r}U$&[\%]                          &   4.06       &   1.44       &   -0.04      &   -1.06      &   -1.77      &   -2.27      \\
\hline
\hline
\multirow{8}{*}{\parbox{0.8mm}{r\\o\\t\\a\\t\\i\\o\\n}} &
                   $\omega^{*}_\text{LBM}$ & $[\si{1\per\second}]$  &  1.97        &   1.18       &   0.829      &   0.638      &  0.519       &   0.436      \\
               &   $\delta_\omega$&[\%]                             &  8.16        &   4.78       &   5.12       &   4.02       &  3.17        &   2.77       \\
               &   $\textup{Re}_{\text{tip},d}$&                      &  0.013       &   0.011      &   0.011      &   0.010      &  0.010       &   0.010      \\
               &   $\textup{Re}_{\text{tip},L}$&                  &  0.025       &   0.034      &   0.042      &   0.051      &  0.060       &   0.068      \\
\cline{2-9}
               & $\omega_\text{Tir,nC}$&$[\si{\per\second}]$         &  4.69        &  1.99        &   1.25       &   0.897      &  0.692       &   0.561      \\
               & $\Delta_\text{r}\omega$&[\%]                       &  -57.9       &  -40.7       &   -33.8      &   -28.8      &  -25.0       &   -22.2      \\
\cline{2-9}
               & $\omega_\text{Tir,wC}$&$[\si{\per\second}]$         &  1.36        &  0.909       &   0.682      &   0.545      &  0.455       &   0.390      \\
               & $\Delta_\text{r}\omega$&[\%]                       &  44.8         &  29.7       &   21.6       &   17.0       &  14.1        &   12.0       \\
\hline
   \end{tabular}
}
\end{table}

\end{appendix}

% Acknowledgements
\section*{Acknowledgements}
The authors would like to thank % `Hier k\"onnte Dein Name stehen'
Gaby Fleig for support in the correction process.
The first author is grateful to the RRZE and LRZ for providing the computational resources on LiMa and SuperMUC, respectively.
The second author would like to thank the Deutsche Forschungsgemeinschaft (DFG) for partially funding this project through the Cluster of Excellence `Engineering of Advanced Materials' in Erlangen.
The fourth author gratefully acknowledges support by the Institute of Mathematical Sciences at the National University of Singapore, where part of this work was performed.

% \section*{References}
\bibliographystyle{elsarticle-num}
\bibliography{bibfileTR}
% \input{tumbling_ecrc.bbl}
%\bibliography{tumbling_ecrc}

\end{document}